\documentclass{aa}  
\usepackage{graphicx,amsmath}
\usepackage{txfonts}
\begin{document}
\newcommand{\mearth}{M_\oplus}
\newcommand{\tacc}{t_{acc}}
\newcommand{\tdyn}{t_{dyn}}
   \title{On the formation and  migration of giant planets in circumbinary discs}


   \author{A. Pierens
          \and
          R.P Nelson
          }

   \offprints{A. Pierens}

   \institute{Astronomy Unit, Queen Mary, University of London, Mile End Rd, London, E1 4NS, UK\\
     \email{a.pierens@qmul.ac.uk}       
             }


 
  \abstract
   {}
   {
We present the results of hydrodynamic simulations of the formation and subsequent orbital evolution of giant planets embedded in a circumbinary disc. 
The aim is to examine whether or not giant planets can be found to orbit stably in close binary systems.
}
   {We performed numerical simulations using a grid--based
   hydrodynamics code. We assume that a $20\;\mearth$ core has migrated to the edge of the inner cavity formed by the binary where it remains trapped by corotation
torques. This core is then allowed to accrete gas from the disc,
 and we study its orbital evolution as it grows in mass. 
For each of the two accretion time scales we considered, 
we performed three simulations. In two of the three simulations, 
we stop the accretion onto the planet once its mass becomes 
characteristic of that of Saturn or Jupiter. In the remaining case, 
the planet can accrete disc material freely in such a way that its 
mass becomes higher than Jupiter's. }
   { The simulations show different outcomes
   depending on the final mass $m_p$ of the giant.  
For $m_p=1\;M_S$ (where $M_S$ is Saturn's mass), we find that the planet 
migrates inward through its interaction with the disc until its 
eccentricity becomes high enough to induce a torque reversal. 
The planet then migrates outward, and the system remains
stable on long time scales. For $m_p \ge 1 M_J$ (where $M_J$ is 
Jupiter's mass) we observed two different outcomes.
In each case the planet enters the 4:1 resonance with the 
binary, and resonant interaction drives up the eccentricity of the planet 
until it undergoes a close encounter with the secondary star, leading to 
scattering. The result can either be ejection from the system
or scattering out into the disc followed by a prolonged period of
outward migration. These results suggest that 
circumbinary planets are more likely to be quite common in
the Saturn-mass range. Jupiter--mass circumbinary planets
are likely to be less common because of their less stable evolution,
but if present are likely to orbit at large distances from the 
central binary.}
   {}


   \maketitle
%

\section{Introduction}

Among the approximately 260 extrasolar planets known at the time of writing
  (e.g.  {\tt
  http://exoplanet.eu/}), about 30 reside in binary or multiple-star
 systems, with  most of them orbiting one stellar component in
  so-called S-type orbits (Eggenberger et al. 2004; Mugrauer et al. 
  2007). The majority of binary stars hosting planets have  orbital
  separation $a_b \ge$ 100 AU. However, there are a few cases of short
  period binary systems with $a_b \sim$ 20 AU like Gliese 86, $\gamma$
  Cephei and HD 41004 in which planets have been discovered to orbit at
  1-2 AU from the primary (Eggenberger et al. 2004, Mugrauer \&
  Neuhauser 2005). As suggested by studies
  of the long-term stability of planets in binary
  systems (e.g. Holman \& Wiegert 1999), close binaries with $a_b\sim$
  1 AU can, in principle, harbour planets evolving on a P-type orbit which encircles
  the two components of the binary system. One such circumbinary planet with
  mass of  $m_p=2.5\;M_J$ has been detected orbiting at 23 AU from the
  radio pulsar binary PSR 1620-26. Another with mass of
  $m_p=2.44\;M_J$ was found evolving around a system which consists of the star
  HD 202206 and its 17.4 $M_J$ brown dwarf companion (Udry et al. 2002). Because short period binaries are
  often rejected from observational surveys, circumbinary planets
  have not yet been observed in binary systems composed of two main
  sequence stars. \\
  Several circumbinary discs, however, have been
  detected around spectroscopic binaries like DQ Tau, AK Sco and GW
  Ori. In GG Tau, the circumbinary disc has been directly imaged and
  has revealed the presence of an inner disc cavity due to the tidal
  torques exerted by the central binary (Dutrey et al. 1994). The
  existence of these circumbinary discs combined with the fact that
  $\sim$ 50 \% of solar-type stars are members of binaries (Duquennoy
  \& Mayor 1991) suggests that  circumbinary planets may be
  common provided that planet formation can occur inside such discs.\\

To date, several theoretical studies focused on planet formation in
close binary systems indicate that planetesimal accretion should be
possible within circumbinary discs. Moriwaki \& Nakagawa (2004) found
that planetesimals can grow in gas-free circumbinary discs only in
regions farther out that $\sim$ 13 AU from a binary with orbital separation $a_b=1$ AU, eccentricity 
$e_b=0.1$ and mass ratio $q_b=0.2$. The influence of gas drag was studied
recently by Scholl et al. (2007) who showed that for a binary with the same
parameters, the eccentricity damping provided by the disc can enable
planetesimal accretion to occur in regions located below $\sim4$ AU from the central
binary. The later stages of planet formation in which Earth-mass
planets form by accumulation of embryos was investigated by Quintana
\& Lissauer (2006). These authors found that planetary systems similar to those around
single stars can be formed around binaries, provided that the ratio
of the binary apocentre distance to planetary orbit is
$\le 0.2$. In general binaries with larger maximum separations lead to
planetary systems with fewer planets.\\
Recently, the evolution of
Earth-mass bodies embedded in a circumbinary disc and undergoing
type I migration because of disc torques (e.g. Ward 1997) was
examined by Pierens \& Nelson (2007) (hereafter referred to as Paper I). In this work,
it was found that the inward drift of a protoplanet can be stopped near
the edge of the cavity formed by the binary. Such an effect arises
because in this region, the
gradient of the disc surface density is such that the
planet experiences strong positive corotation torques which can
eventually counterbalance the negative differential Lindblad torque,
thereby leading to the halting of migration (Masset et al. 2006). In a
subsequent paper, Pierens \&
Nelson (2008a) extended this work by investigating the issue of how multiple protoplanets
interact with each other if they form at large distance from the
binary and successively migrate toward the cavity edge. The
simulations performed by Pierens \& Nelson (2008a)  of pairs of planets interacting with each other indicated
different outcomes such as resonant trapping or
orbital exchange, depending on the ratio between the masses of the 
planets. Interestingly, in simulations involving more than two
planets, planetary growth resulting from  scattering and collisions  between
protoplanets was found to be feasible. This implies that giant cores
might be formed in circumbinary discs, resulting eventually in a gas
giant planet orbiting near the cavity edge.\\
The orbital evolution of Jovian mass planets embedded in circumbinary
discs was studied by Nelson (2003). This work showed that giant
planets undergoing type II migration (e.g. Lin \& Papaloizou 1993; Nelson 2000)
are likely to enter the 4:1 resonance with the
binary. In that case, the subsequent evolution of the giant can be twofold,
depending on whether or not the resonance is stable. For systems in which
the 4:1 resonance is stable, the planet remains near or at
the resonance. However, it appears that there is a finite probability
for the system to be unstable, in which case the planet can be ejected
from the system due to close encounters with the binary.\\

In this paper, we extend the work of Nelson (2003) by studying the
whole evolution of a circumbinary planet during its growth from a core into
a gas giant. To address this issue, we consider a scenario in
which a 20 $\mearth$ core initially trapped at the edge of the cavity
can slowly
accrete gas from the disc. As the planet grows, the onset of
non-linear effects as well as gap formation can exclude gas material
from the coorbital region, thereby cancelling the effects of
corotation torques. This can subsequently lead to the planet migrating inward again,
until it becomes trapped eventually in a mean motion resonance with
the binary. Here, we present the results of hydrodynamic calculations
aimed at simulating such a scenario. In particular, we want to examine
how the final outcome of the system depends on the accretion rate onto
the planet as well as on the final mass of the giant. Interestingly,
the results of the simulations indicate that only Saturn-mass giant planets can evolve stably
in circumbinary discs. Most of the calculations of embedded giants
with masses of $m_p \ge 1\;M_J$ resulted in  close
encounters between the planet and the binary, leading eventually to the planet
being completely ejected from the system.\\

This paper is organized as follows. In Section 2, we describe the
hydrodynamical model. The results of the simulations are discussed in
Section 3. We finally summarise and present our conclusions in Section 4.


\section{Hydrodynamical model}

\subsection{Numerical method}
We consider a 2D disc model in which all physical quantities are
vertically averaged and we work in polar coordinates $(r,\phi)$ with the
origin located at the centre of mass of the binary. The equations governing the disc evolution as
well as the equations of motion for the binary plus planet system can
be found in Paper I. The  equations for the disc are solved using the hydrocode
Genesis, which has been tested extensively against other codes
(De Val-Borro et al. 2006), and employs a second order
numerical scheme based on the monotonic transport algorithm (Van Leer
1977). Included in this code is a fifth-order Runge-Kutta integrator (Press et al. 1992) 
 used to compute the evolution of the planet and binary orbits.\\
As in Paper I, we use $N_r=256$ radial grid cells uniformly
distributed between $r_{in}=0.5$ and $r_{out}=6$ and $N_{\phi}=380$
azimuthal grid cells. We adopt also the same computational units in
which the mass of the binary is $M_\star=1$, the gravitational
constant is $G=1$ and the radius $r=2$ in the computational domain
corresponds to 5 AU. The unit of time is
$\Omega^{-1}=\sqrt{GM_*/a_b^3}$, where $a_b=0.4$ is the initial binary
separation. In the following, we report our results in units of the
initial orbital period of the binary $P=2\pi\Omega^{-1} $.\\
In this paper, the planet is allowed to accrete gas from the
disc.  Accretion by the protoplanet can be modelled
by removing at each time-step a fraction of the gas 
located inside the Roche lobe of the planet and then adding the
corresponding amount of matter to the mass of the planet (e.g. Kley
1999; Nelson et al. 2000). Here, the removal rate is chosen such that
the accretion time scale onto the planet is $t_{acc}=t_{dyn}$, where
$t_{dyn}$ is the orbital period of the planet. This corresponds to the
maximum rate at which the planet can accrete gas material (Kley
1999). In order to examine how the final outcome
of the system depends on the accretion rate, we have also performed
simulations with $t_{acc}=10\;t_{dyn}$.

When calculating the gravitational force between the disc and planet
we adopt a gravitational softening parameter $b=0.6H$, where $H$ is the disc
height at the planet location. Material contained within the planet Hill sphere
does not contribute to the gravitational acceleration experienced by the planet.

\subsection{Initial conditions}

\begin{figure}
   \centering
   \includegraphics[width=0.95\columnwidth]{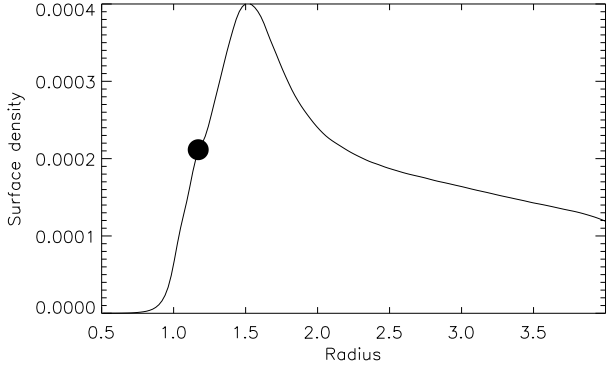}
      \caption{This figure shows the initial disc surface density
      profile. The initial position of the planet with mass of
      $m_p=20\;\mearth$ is represented by a black circle.}
         \label{rho0}
   \end{figure}

In Paper I, we presented the results of simulations of protoplanets
with masses of $m_p=$ 5, 10 and 20 $\mearth$ embedded in circumbinary
discs. We found that in each case, the inward migration of the
protoplanet is halted due to the action of corotation torques which
operate strongly at the cavity edge. Here, we extend the model 
presented in this earlier paper and examine how such a trapped
protoplanet evolves as it accretes gas from the disc and 
grows to become a giant planet.  In order
to investigate this issue, we restarted one of the simulations presented in 
paper I for which  $m_p=20\;\mearth$ at a point in time when the
planet is trapped at the cavity edge. Therefore, in this new series of
simulations, the protoplanet evolves initially on an orbit with $a_p\sim 1.2$
and $e_p\sim 0.02$, which are the values for the 
semi-major axis and eccentricity that a 20 $\mearth$ body finally attains once
its migration has been stopped (see Paper I).\\
The initial semi-major axis and eccentricity of the binary are
$a_b\sim 0.39$ and $e_b\sim 0.08$ respectively. These values
correspond to the ones that an initially circular binary 
with $a_b=0.4$ and mass ratio
$q_b=0.1$ ultimately attains as it interacts with a circumbinary disc. In Paper I,
we indeed showed that the evolution outcome of such a system is an
equilibrium configuration for which the disc structure as well as the
binary eccentricity remain unchanged. Interestingly, from the time
this quasi-steady state is reached, we found that the apsidal 
lines of the disc and binary are almost aligned. \\

The disc model used in this work is the same as that of Paper
I. Accordingly, the  aspect ratio is constant and equal to $H/r=0.05$.  The initial disc surface density
profile is presented in Fig. \ref{rho0}. For numerical reasons (see Section \ref{sec:bc}), we use a 
low-density region from $r=4$ to $r=6$ . From $r\sim 2$ to $r=4$, the surface
density is $\Sigma(r)=\Sigma_0\;r^{-1/2}$ where $\Sigma_0$ is chosen
such that the unperturbed disc would contain 0.01 $M_\odot$ inside 10 AU (we
assume that the initial binary separation corresponds to 1 AU). We further 
note that such a density profile corresponds to the equilibrium density 
for an accretion disc with a constant kinematic viscosity (e.g. Gunther \& 
Kley 2002). In the
inner parts, the disc presents an inner cavity which is created
self-consistently from simulations of binary-disc interactions (see Paper I). It is worthwhile to notice that here, the onset of
non-linear effects due to the presence of
a 20 $\mearth$ initially located at $r\sim 1.2$ can modify slightly
the gap profile.\\
We model the disc turbulent  viscosity using the
``alpha'' prescription for the effective kinematic viscosity
$\nu=\alpha c_s H$ (Shakura \& Sunyaev 1973). In Paper I, we set
$\alpha=10^{-4}$ because using larger values caused the binary
separation to decrease too rapidly to allow an equilibrium
configuration to be obtained (see Paper I for details). In this work
however, we examine the evolution of giant planets undergoing type II migration. Because in that case the migration rate
of the planet is controlled by the disc viscous evolution, we decided here
to use a more realistic $\alpha$ value, which probably lies in the range
$10^{-3}-10^{-2}$ in real circumstellar dics. In order to obtain a
disc model in which $\alpha=10^{-3}$, the calculations were started
with  $\alpha$ increasing slightly
 from $10^{-4}$ to the desired value in $\sim1500$ binary
orbits. Although not sufficient for a new equilibrium
configuration to be established, this gives a sufficient time to the binary plus disc system to
readjust to the new $\alpha$ value.

\subsection{Boundary conditions}\label{sec:bc}
In order to avoid any wave reflection at the outer edge of the
computational domain, we impose a low-density region between $r=4$ and
$r=6$ using a taper function.\\
At the inner boundary, we model the accretion onto the central star by
 setting the radial velocity in the inner ghost zones to $v_r=\beta
 v_r(r_{in})$, where $v_r(r_{in})=-3\nu/2r_{in}$ is the gas drift
 velocity at the disc inner edge due to viscous evolution and where $\beta$
 is a free parameter. Following Pierens \& Nelson (2008b), we set
 $\beta=5$ in the simulations presented here.

\section{Results}

\begin{figure}
   \centering
   \includegraphics[width=0.95\columnwidth]{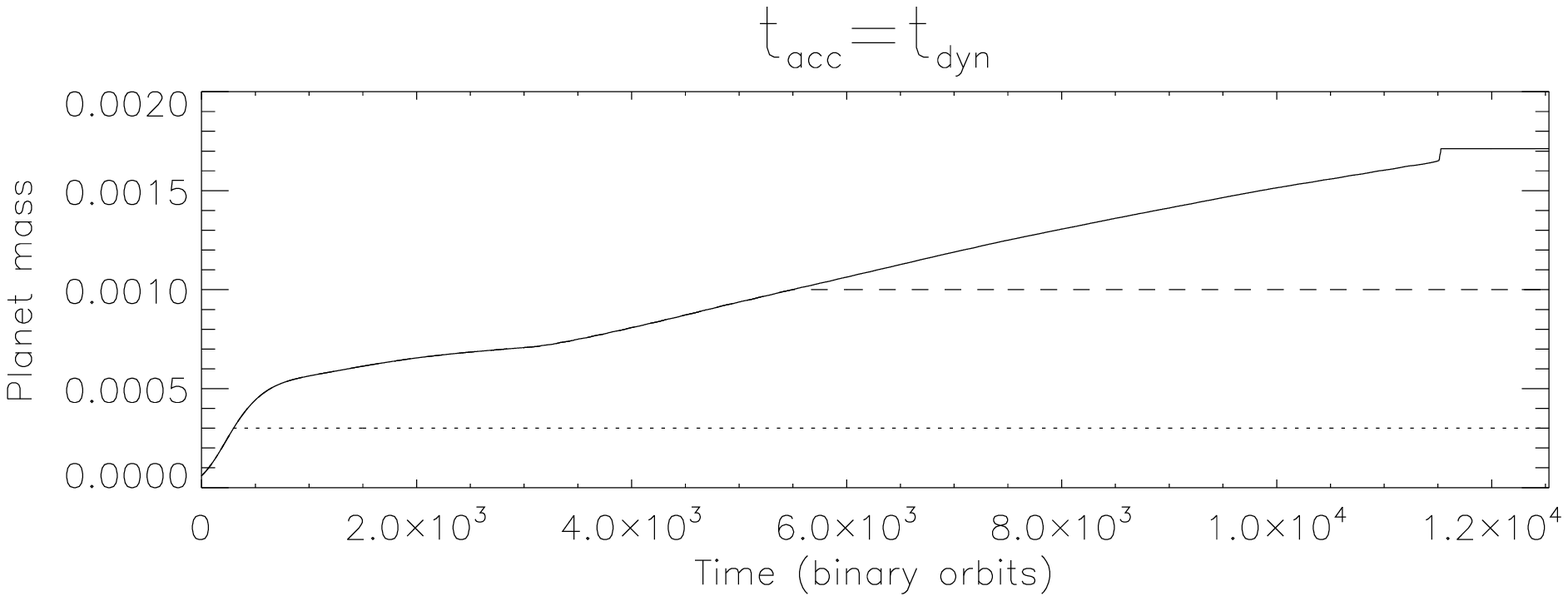}
   \includegraphics[width=0.95\columnwidth]{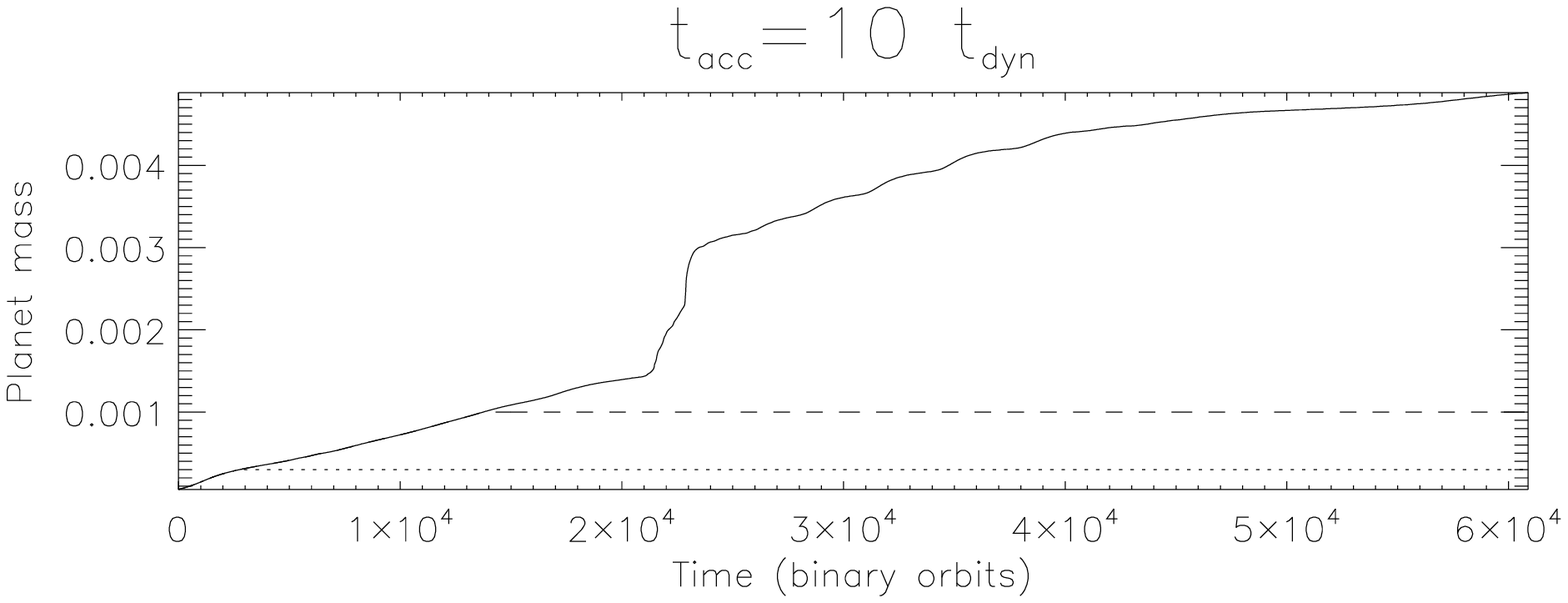}
      \caption{This figure shows the mass of the planet as a function
      of time for simulations in which $\tacc=\tdyn$ ({\it upper
      panel}) and $\tacc=10\;\tdyn$ ({\it lower
      panel}). The dotted  (resp. dashed) line corresponds to
      simulations in which the final mass of the planet is 1 $M_s$
      (resp. 1 $M_J$). In  simulations represented by the solid line, 
      the final mass of the planet is not imposed.}
         \label{mass}
   \end{figure}

\begin{figure*}
   \centering
   \includegraphics[width=\columnwidth]{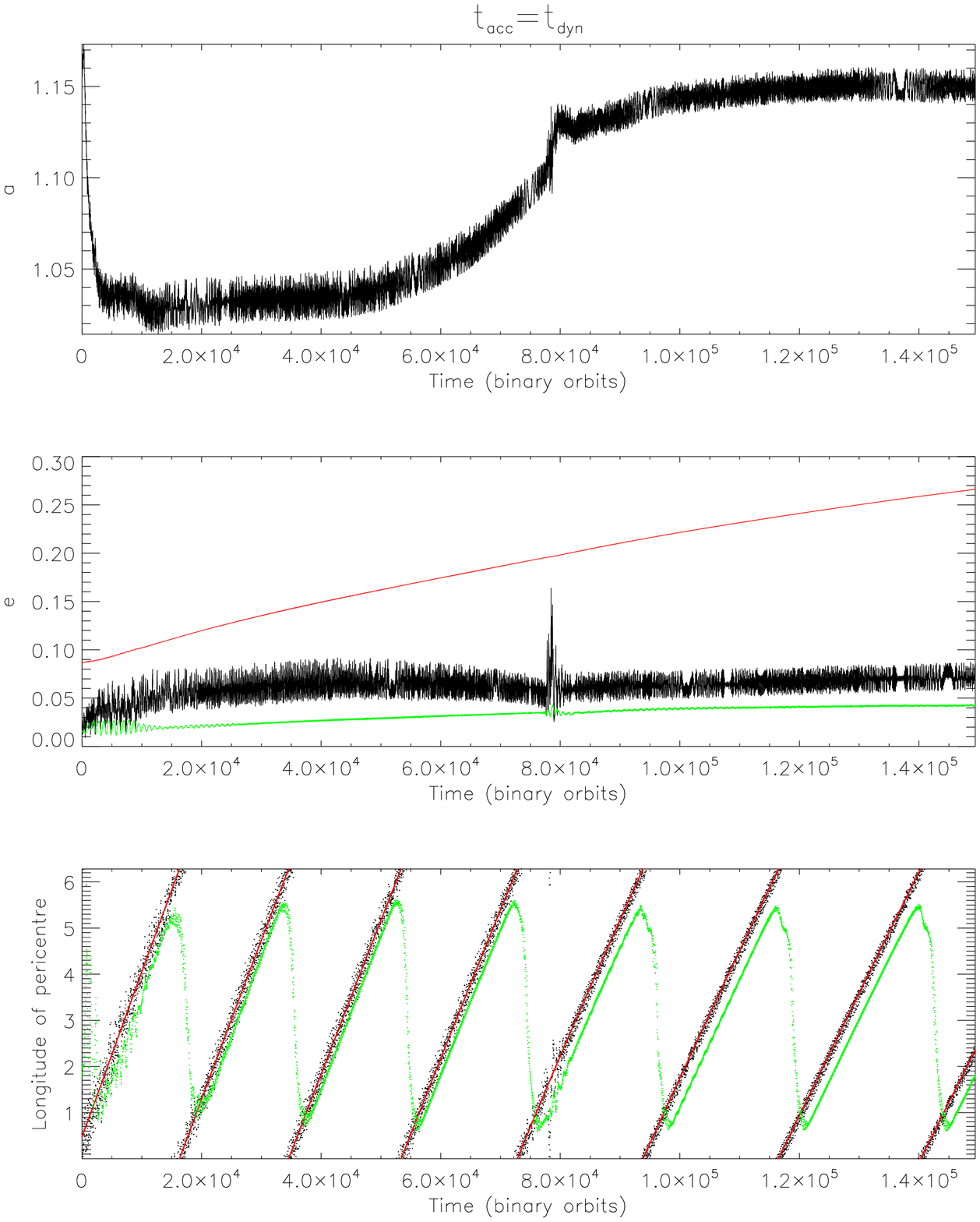}
   \includegraphics[width=\columnwidth]{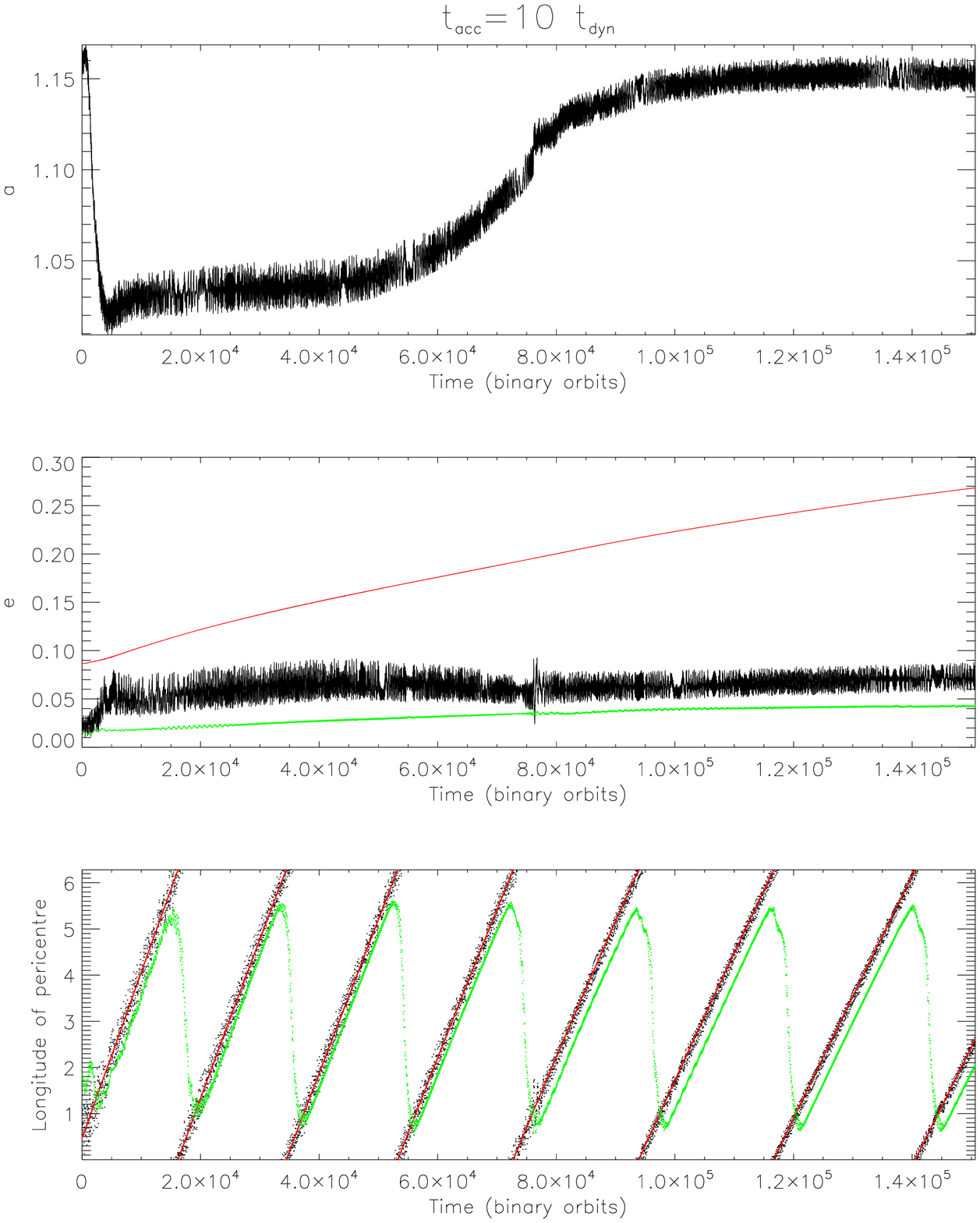}
      \caption{This figure shows the evolution of the system for
      models in which $m_p=1\;M_S$. Here, the left and right panels correspond
      to $\tacc=\tdyn$ and $\tacc=10\;\tdyn$ respectively. {\it Top:}
      evolution of the semi-major axis of the planet. {\it Middle:}
      evolution of the eccentricities for the planet (black),
      binary (red) and disc (green). {\it Bottom:} evolution of the
      longitudes of pericentre for the planet (black), binary (red)
      and disc (green).}
         \label{saturn}
   \end{figure*}

For each value of the accretion rate onto the planet we consider, we
have performed three simulations which differ in the final mass
attained by the giant. In the first and second calculations, we prevent
further growth of the planet once its mass becomes $m_p=1\;M_s$ (where
$M_s$ is Saturn's mass) and $m_p=1\;M_J$ respectively. In the last run
however, the final mass of the planet is not restricted and we allow
the latter to accrete gas freely from the disc. The mass of
the planet as a function of time is presented in Fig. \ref{mass}
for each run.

\subsection{Evolution of Saturn-mass planets}

For both values of $\tacc$,  the semi-major axis evolution  of a 20 $\mearth$ body which grows to become a Saturn-mass planet is displayed in the
upper panel of Fig. \ref{saturn}. In the model with $\tacc=\tdyn$, the
planet mass reaches $m_p=1\;M_s$ at $t\sim 300$ P, and this mass is
attained  at $t\sim 2800$ P in the model with $\tacc=10\;\tdyn$ (see
Fig. \ref{mass}). The orbital evolution of the planet is, however, very
similar in both cases and typically proceeds as follows. At the
beginning of the simulation, gap formation due to the growth of the
protoplanet can exclude material from the coorbital
region, which results in the (negative) differential Lindblad 
torques being no longer
couterbalanced by the (positive) corotation torques. 
The planet thus migrates inward again. As it migrates,
the interaction with the central binary makes $e_p$ 
increase slowly, as observed in the middle panel of Fig. \ref{saturn}
which shows the evolution of the eccentricities of the binary, disc
and planet for both values of $\tacc$. 
We define the disc eccentricity $e_d$ by:
\begin{equation}
e_d=\frac{\int_0^{2\pi}\int_{r_{in}}^{r_{max}}\Sigma\; e_c dS}{\int_0^{2\pi}\int_{r_{in}}^{r_{max}}\Sigma\; dS};
\end{equation}
where  $r_{max}$ is set to $r_{max}=3$ and where $e_c$ is the eccentricity of the disc computed at the centre of each grid
cell. This orbital element can be deduced by computing the eccentricity vector
${\bf e}=({\bf v}\times{\bf h})/M_\star-{\bf r}/r$, where ${\bf v}$ and ${\bf h}$ are respectively the velocity and angular momentum of the disc at the cell centre. \\

\begin{figure}
   \centering
   \includegraphics[width=0.95\columnwidth]{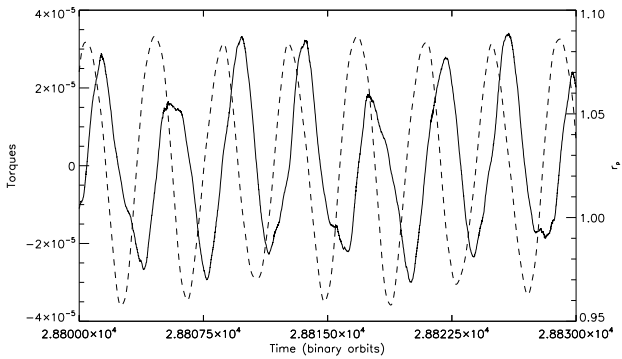}
      \caption{This figure shows, over a few planetary orbital periods, 
       the evolution of the torques exerted
      by the disc on the planet (solid line) as well as the  time evoution
      of the orbital the position of
      the planet $r_p$ (dashed line).}
         \label{zoom}
   \end{figure}

Once the planet eccentricity reaches  $e_p\sim 0.05$, which occurs at
$t\sim 500$ P,  the migration of the planet stops and
then reverses. Previous work of Papaloizou \& Larwood (2000) indicated
that an eccentric low-mass protoplanet can experience net positive disc torques when
$e_p\sim 1.1 H/R$, owing to the latter
crossing resonances in the disc that do not overlap the orbit at
low-eccentricities. As pointed out by these authors, this occurs because a planet evolving on
a high eccentric orbit rotates
more slowly than the disc at apocentre, which results in
the outer disc exerting a positive torque on the planet. Although here
the planet mass is in the Saturnian mass range, it appears that
the observed migration reversal is caused by  a similar phenomenon. In
order to demonstrate that such an effect is at work here, we have
plotted in Fig. \ref{zoom} the evolution of both the disc torques and
 planet orbital position $r_p$ over a few orbital periods of the
planet. In agreement with Papaloizou \& Larwood (2000), the disc
torques exerted on the planet are positive (negative) when the latter is at
apocentre (pericentre). Clearly, there is a slight imbalance 
between the positive and
negative torques, due to the planet orbiting at a cavity edge,
which combined with the fact that the planet spends
more time at apocentre, favours a net positive torque on the planet
and outward migration. We note that in
Fig. \ref{zoom}, there is a slight phase shift between the two curves which
corresponds to the time needed for the planet to create an inner
(outer) wake at apocentre (pericentre).\\

\begin{figure*}
   \centering
   \includegraphics[width=6cm]{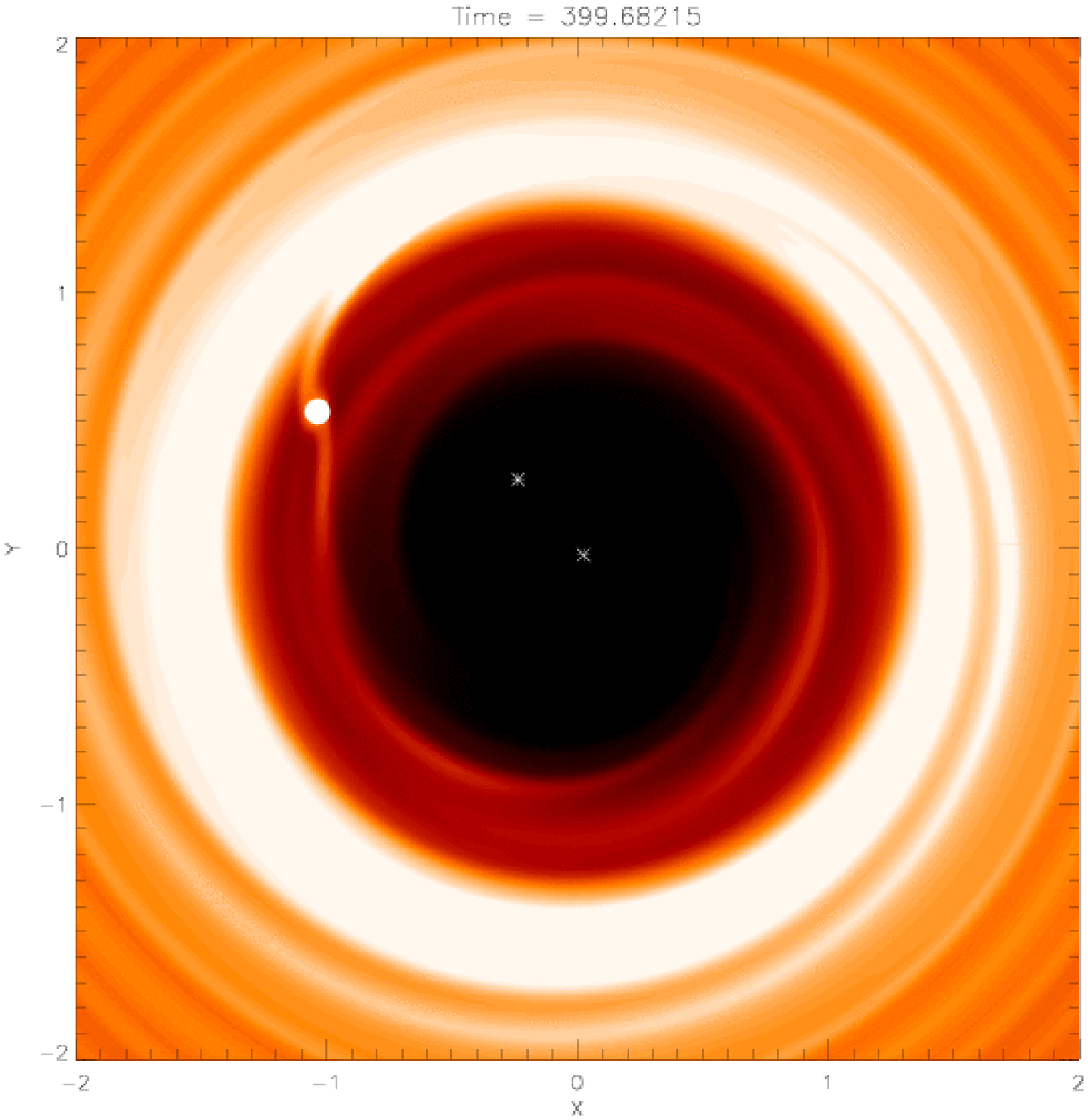}
    \includegraphics[width=6cm]{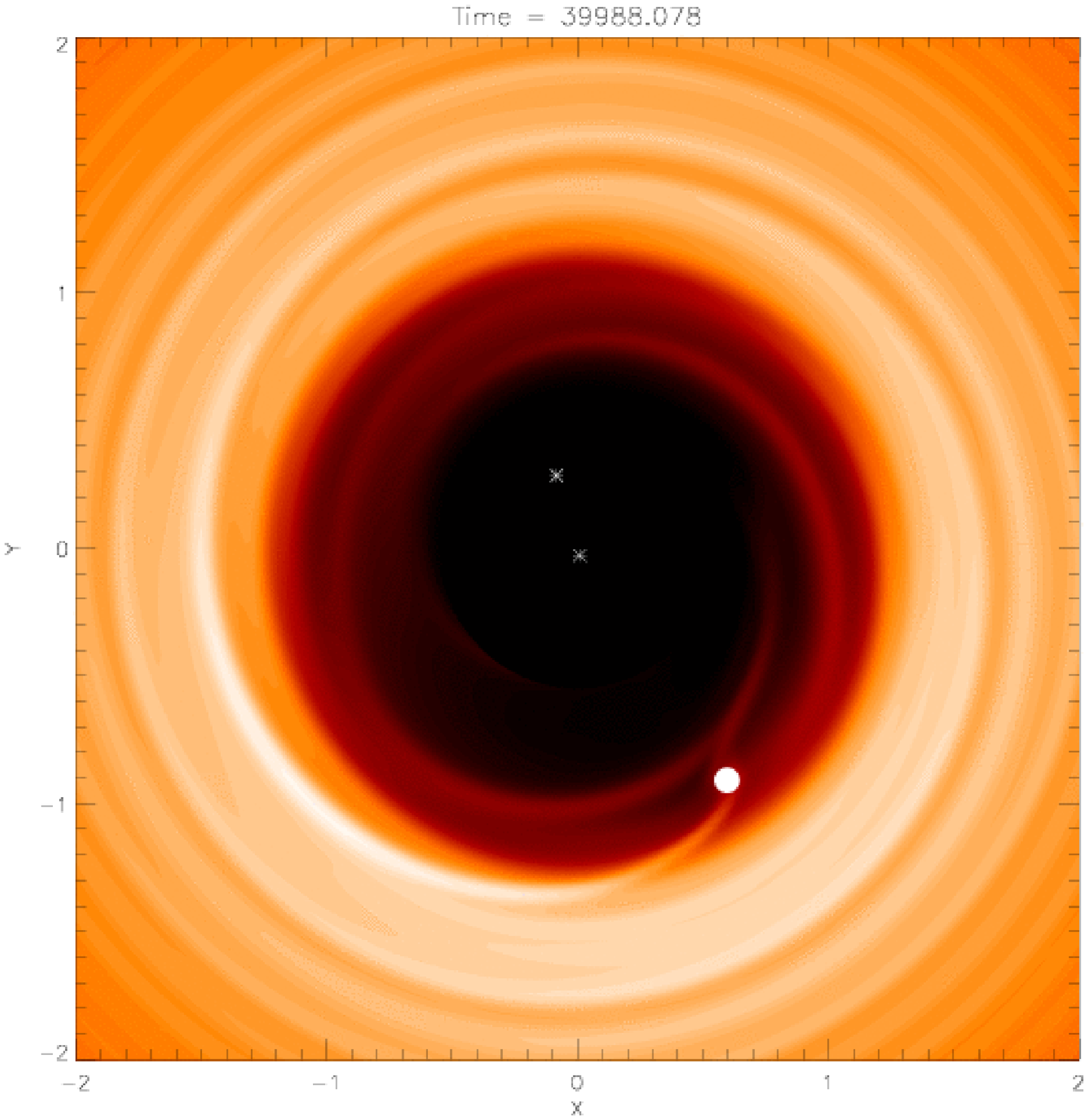}
     \includegraphics[width=6cm]{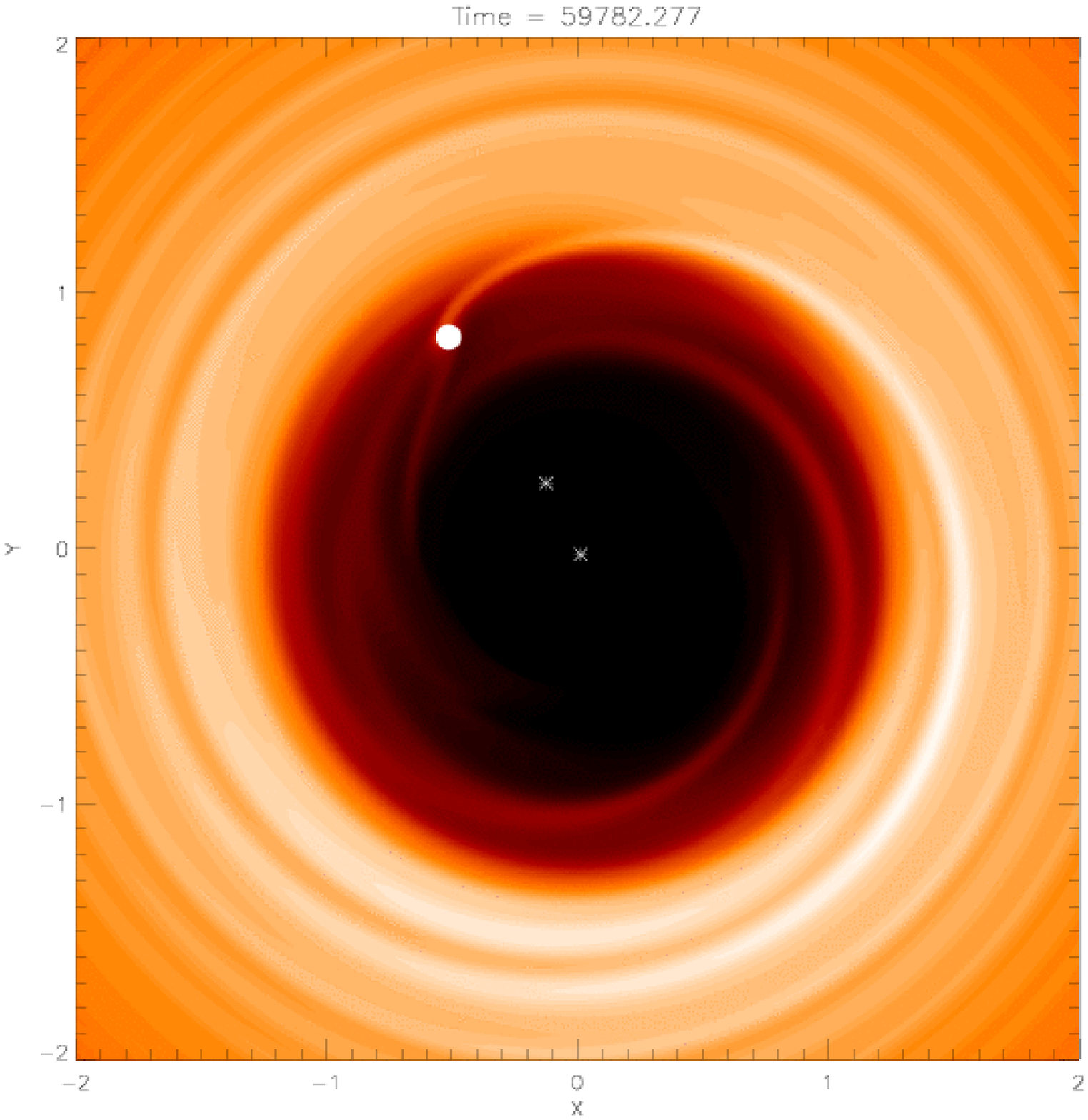}
     \includegraphics[width=6cm]{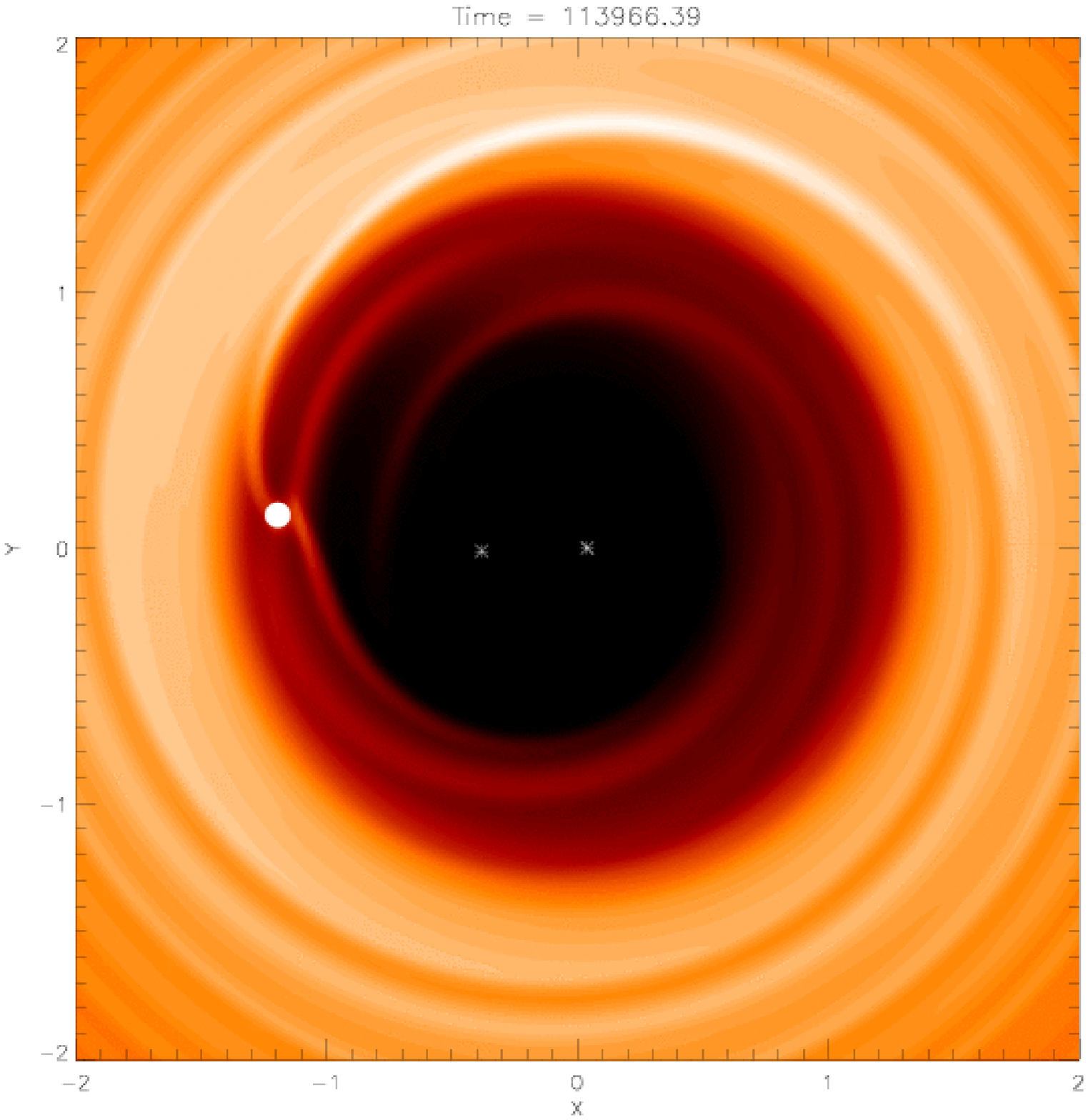}
      \caption{This figure shows, for the model in which $m_p=1\;M_S$
      and $\tacc=\tdyn$,  snapshots of the disc surface density
      at times shown above the plots. Here, the planet is
      represented by a white circle.}
         \label{saturn2d}
   \end{figure*}

Whereas this reversed migration proceeds very slowly until $t\sim
4.5\times 10^4$ P, subsequent evolution involves exponential growth of
the planet migration rate, which is characteristic of an episode of
runaway migration (Masset \& Papaloizou 2003). Saturn-mass planets are
known to be good candidates for such a
migration regime in massive
protoplanetary discs,  due to their ability to create a coorbital mass deficit larger than
their own  mass. As noticed by Masset \& Papaloizou (2003), outward runaway
migration can eventually occur in discs with shallow surface
density profiles, provided that the planet initially migrates with a
significantly positive drift rate. Here, two effects contribute to
make such a phenomenon possible. First, prior to the
period of runaway migration, the planet migrates outward over 
$\sim 2\times 10^4$ binary orbits, which is clearly larger than the
planet libration timescale. Second, the planet evolves in a region
of strong positive surface density gradient such that the coorbital
mass deficit increases as the planet migrates outward. This can be
seen for example in Fig. \ref{saturn2d} which shows, for the model in
which $\tacc=\tdyn$, a series of surface density plots at different
times.\\
Interestingly, this period of outward runaway migration breaks at
$t\sim 8\times 10^4$ P, which corresponds to the time when the planet
passes through the 5:1 resonance with the binary. This can excite the planet
eccentricity to such values that during the course of an orbit, the
radial excursion of the planet can exceed its coorbital width, 
leading to the loss of the
coorbital mass deficit. The fourth panel in Fig. \ref{saturn2d}
displays the disc surface density just after the 5:1 resonance
crossing. Comparing this panel with the second one which corresponds
to a time when the planet is about to undergo runaway migration, it is clear 
that the coorbital region is no longer depleted, which can prevent runaway
migration being sustained. This result is in broad agreement with the
calculations performed by Masset \& Papaloizou
(2003) which indicate a  similar tendency for the runaway
migration to not be maintained.\\

After the 5:1 resonance crossing, continuation of the runs indicates
that the planets undergo slow outward migration for $\sim 5\times
10^4$ binary orbits and then migrate inward again. Due to the very
long run times required by these simulations, the final fate of the
planets remains uncertain. Nevertheless, it seems likely that continued
inward migration will proceed until the eccentricities of the planets are
high enough for the disc torques to become positive, resulting
eventually in the planets migrating outward again. Thus we expect
the long term evolution to consist of periods of inward followed
by outward migration until disc dispersal, resulting in a stable
Saturn--mass circumbinary planet.

The evolution of the longitudes of pericentre of the binary, disc and
planet is depicted in the lower panel of Fig. \ref{saturn}. As in
Paper I, we compute the disc longitude of pericentre according to the
following definition:
\begin{equation}
w_d=\frac{\int_0^{2\pi}\int_{r_{in}}^{r_{max}}\Sigma\; w_c dS}{\int_0^{2\pi}\int_{r_{in}}^{r_{max}}\Sigma\; dS};
\end{equation}
where $w_c$ is the longitude of pericentre of the disc at the centre of
each grid cell. This orbital element is computed from the knowledge of
 the eccentricity vector using the relation $\cos(w_c)=e_x/e$. We note that $w_d$ may be significantly different if the
gravity of the disc was taken into account in the simulations. Indeed, it is well-known that one
of the main effects of self-gravity is to induce prograde precession of
the disc (e.g. Papaloizou 2002). The figure shows that the apsidal line of the planet
is almost perfectly aligned with the ones of the disc and binary. Here,
 it is worthwhile noticing that despite
 the fact that the disc and binary precess at the same rate, the
 calculations show continued growth of the binary eccentricity
 due to resonant interactions between the disc and binary (see middle
 panel of Fig. \ref{saturn}). This
 indicates that the binary plus disc system has not yet reached an
 equilibrium configuration where the eccentricity forcing has saturated.

\subsection{Evolution of Jupiter-mass planets}\label{sec:jup}

\begin{figure*}
   \centering
   \includegraphics[width=\columnwidth]{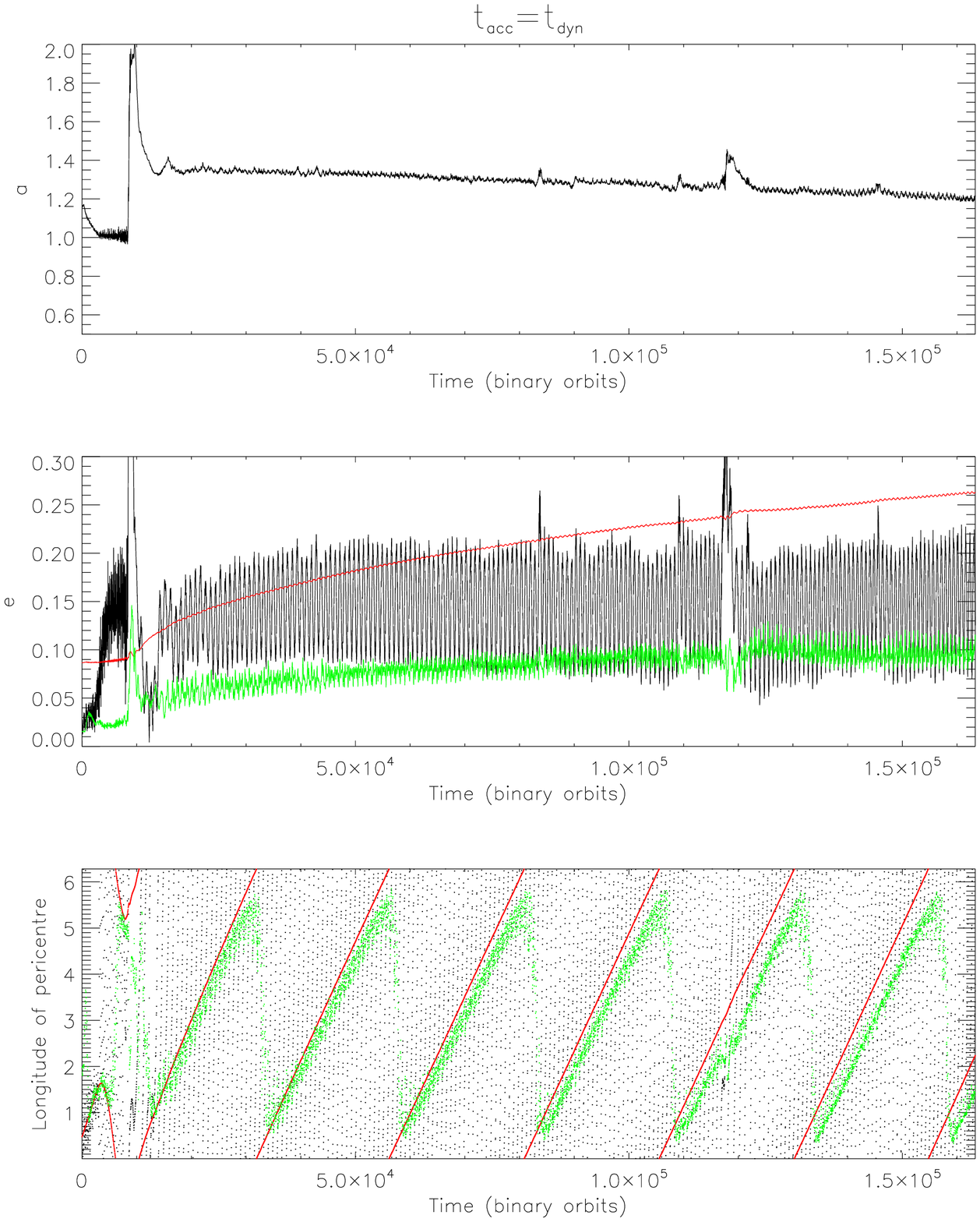}
   \includegraphics[width=\columnwidth]{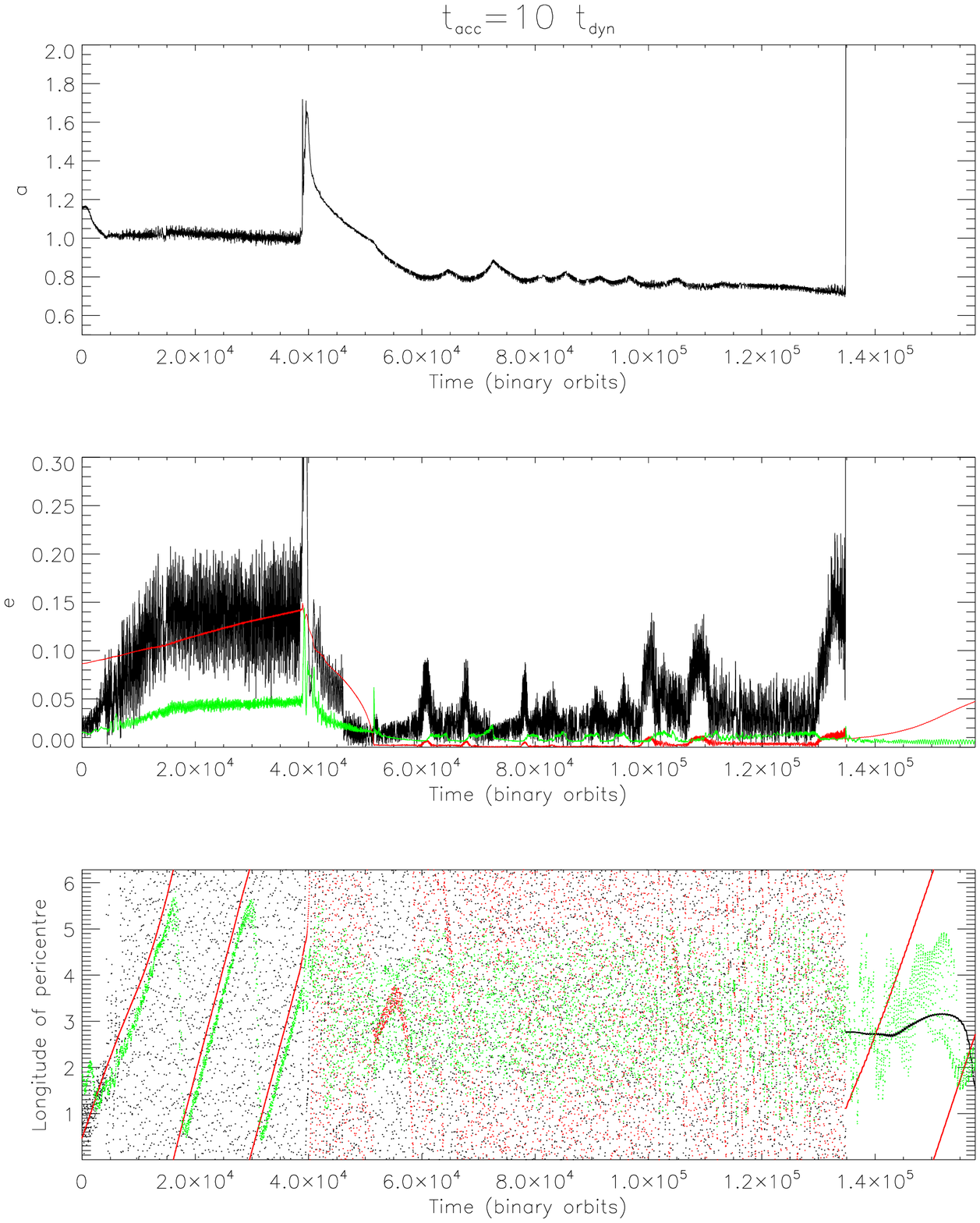}
      \caption{This figure shows the evolution of the system for
      models in which $m_p=1\;M_J$. Here, the left and right panels correspond
      to $\tacc=\tdyn$ and $\tacc=10\;\tdyn$ respectively. {\it Top:}
      evolution of the semi-major axis of the planet. {\it Middle:}
      evolution of the eccentricities for the planet (black),
      binary (red) and disc (green). {\it Bottom:} evolution of the
      longitudes of pericentre for the planet (black), binary (red)
      and disc (green).}
         \label{jupiter}
   \end{figure*}

For the two accretion time scales we consider, the orbital evolution of
a planet with final mass $m_p=1\;M_J$ is illustrated in the upper
panel of Fig. \ref{jupiter}. In the model with $\tacc=\tdyn$ the planet
reaches one Jupiter mass in $\sim 5.5\times 10^3$ binary orbits while it 
reaches this mass in  $\sim 1.4\times 10^4$ orbits in the model with
$\tacc=10\; \tdyn$ (see Fig. \ref{mass}). The early evolution of the planet is 
similar to that described for Saturn-mass planets, involving inward migration of the planet and
continued growth of its eccentricity. However, contrary to  models in 
which $m_p=1\; M_S$, simulations with $m_p=1\;M_J$ 
resulted in the temporary formation of the 4:1 resonance between 
the planet and binary (we note that migration reversal occurs for the planets
with $m_p=1 \; M_S$ before they reach the 4:1 resonance).\\

\begin{figure}
   \centering
   \includegraphics[width=0.95\columnwidth]{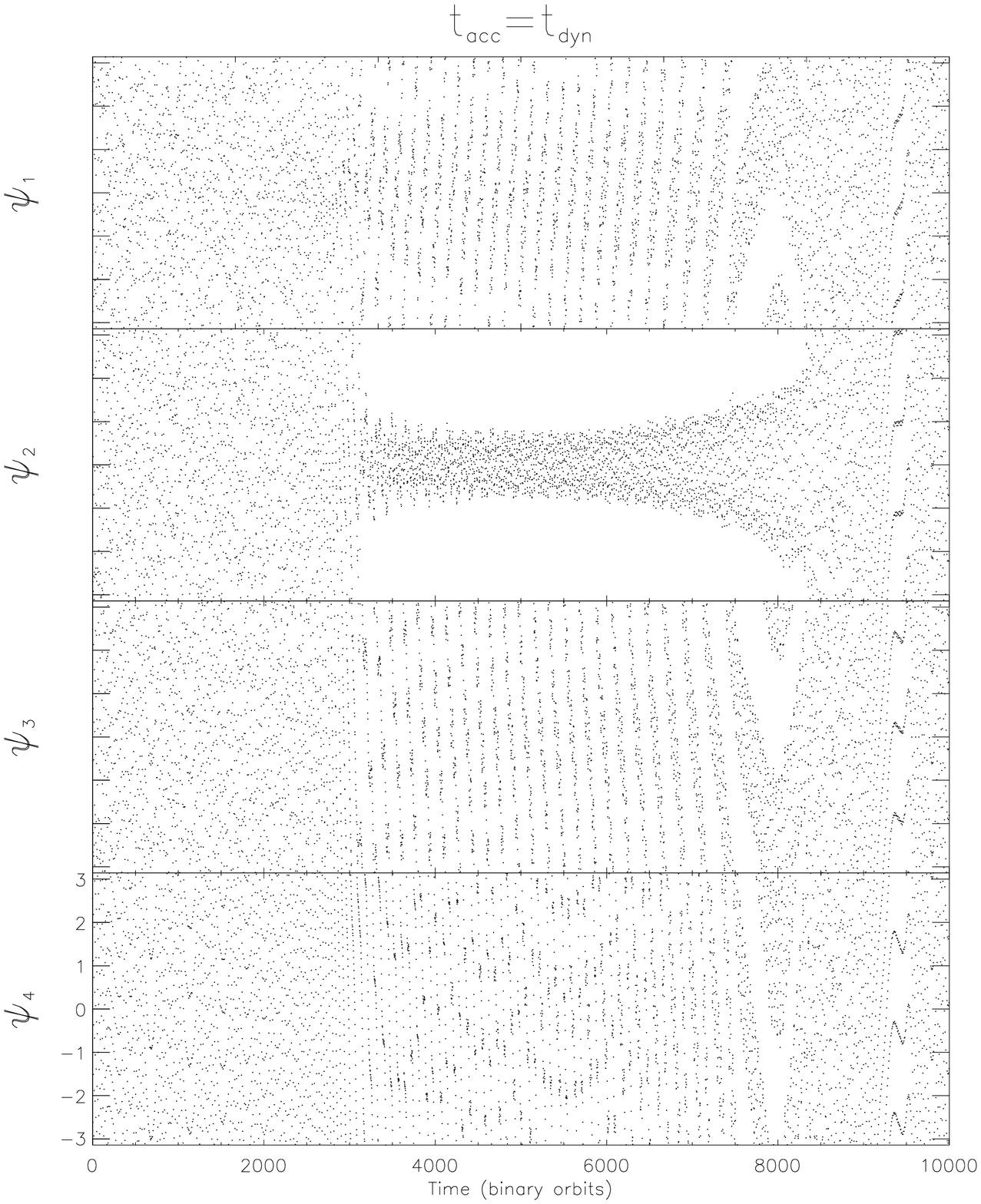}
      \caption{This figure shows, for the simulation with
      $m_p=1\;M_J$ and $\tacc=\tdyn$, the evolution of the resonant angles  $\psi_1$,  $\psi_2$,  $\psi_3$ and
$\psi_4$ associated  with the 4:1 resonance.}
         \label{angles}
   \end{figure}

 For the calculation with $\tacc=\tdyn$, the time
evolution of the resonant angles $\psi_1$,  $\psi_2$,  $\psi_3$ and
$\psi_4$ associated with the 4:1 resonance is
displayed  in Fig. \ref{angles}. These are given by:
\begin{equation}
\begin{array}{l}
\psi_1=4\lambda_b-\lambda_p-3\omega_b,\\
\psi_2=4\lambda_b-\lambda_p-3\omega_p,\\
\psi_3=4\lambda_b-\lambda_p-2\omega_b-\omega_p,\\
\psi_4=4\lambda_b-\lambda_p-2\omega_p-\omega_b,\\
\end{array}
\end{equation}
where $\lambda_b$ ($\lambda_p$) and $\omega_b$ ($\omega_p$) are
respectively the mean longitude and longitude of pericentre of the
binary (planet). Once the resonance is established, which occurs at $t\sim
3\times 10^3\; P$ for this model, the resonant angle $\psi_2$ librates with low
amplitude, thereby indicating that the planet is strongly locked into
the resonance. The evolution of the eccentricities of the planet, disc
and binary is illustrated in the middle panel of
Fig. \ref{jupiter}. As expected, the resonant interaction with the binary is accompanied by
a significant growth of the planet eccentricity $e_p$, which increases up to
$e_p\sim 0.15$. At $t\sim 8.5\times 10^3 \; P$, this eccentricity growth causes the
protoplanet to undergo a close encounter with the binary, leading
subsequently to the planet being extracted from the resonance and
scattered further out in the disc on a high-eccentric orbit with $a_p\sim 2$.\\

A similar outcome is observed in the simulation with  $\tacc=10\;
\tdyn$. Here the planet is scattered at $t\sim 4 \times 10^4\; P$
and evolves subsequently on an orbit with significantly larger 
semimajor axis ($a_p\sim 1.6$) and eccentricity ($e_p\sim 0.3)$. Since a close encounter is generally a chaotic dynamical process, these values are quite different from the ones observed in the simulation with $\tacc=\tdyn$. In fact, these strongly depend on the encounter geometry and therefore may 
significantly differ from one simulation to the other. \\
Prior to this close encounter however,
the evolution of the planet differed noticeably from that corresponding to the model
with $\tacc=\tdyn$. It appears that in the calculation with 
$\tacc=10\; \tdyn $, the
4:1 resonance becomes rapidly 
undefined after its formation, which arises at $t\sim 4\times
10^4$. This is because, compared with the run in which
$\tacc=\tdyn$, the planet mass is significantly lower when the 4:1 
resonance is established, leading to a weaker resonant
locking. Once the resonance is broken, the planet is located just outside
of the 4:1 resonance and migrates slightly outward until 
$t \sim 1.7\times 10^4\; P$. Outward migration is induced by 
positive disc torques due to the planet eccentricity having reached 
$e_p\sim 0.15$ after resonance breaking. Then, the planet 
migrates inward and approaches the 4:1 resonance again, 
but its eccentricity remains sufficiently high for there to be a close
encounter between the planet and central binary system leading to
the planet being scattered by the binary. The planet and binary
undergo multiple encounters during this phase, with significant changes
in their orbital elements occuring (see middle-right 
panel of Fig.~\ref{jupiter}). The planet ends up orbiting interior
to the 4:1 resonance where it migrates inward 
toward the 3:1 resonance. \\

\begin{figure*}
   \centering
   \includegraphics[width=6cm]{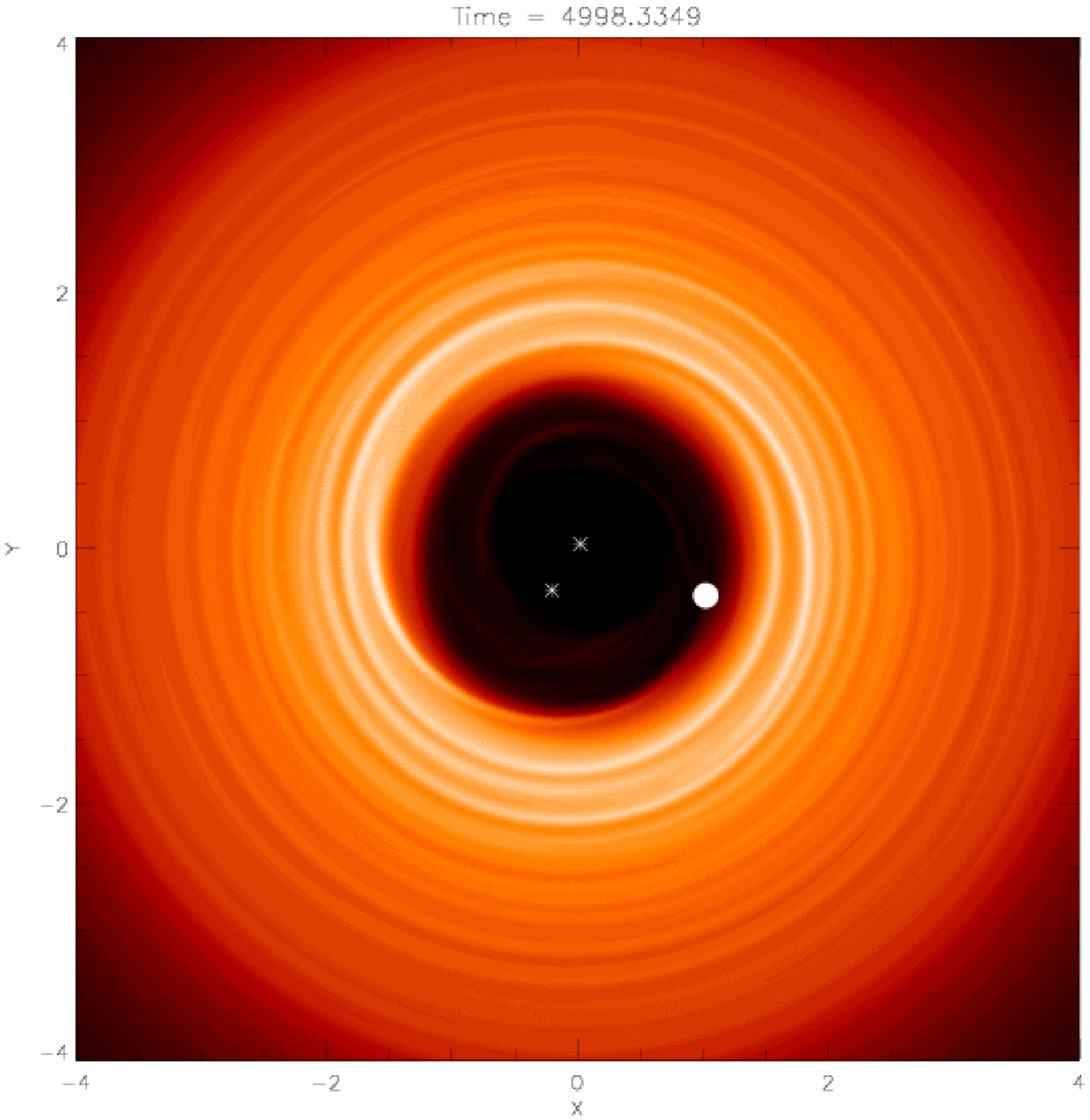}
    \includegraphics[width=6cm]{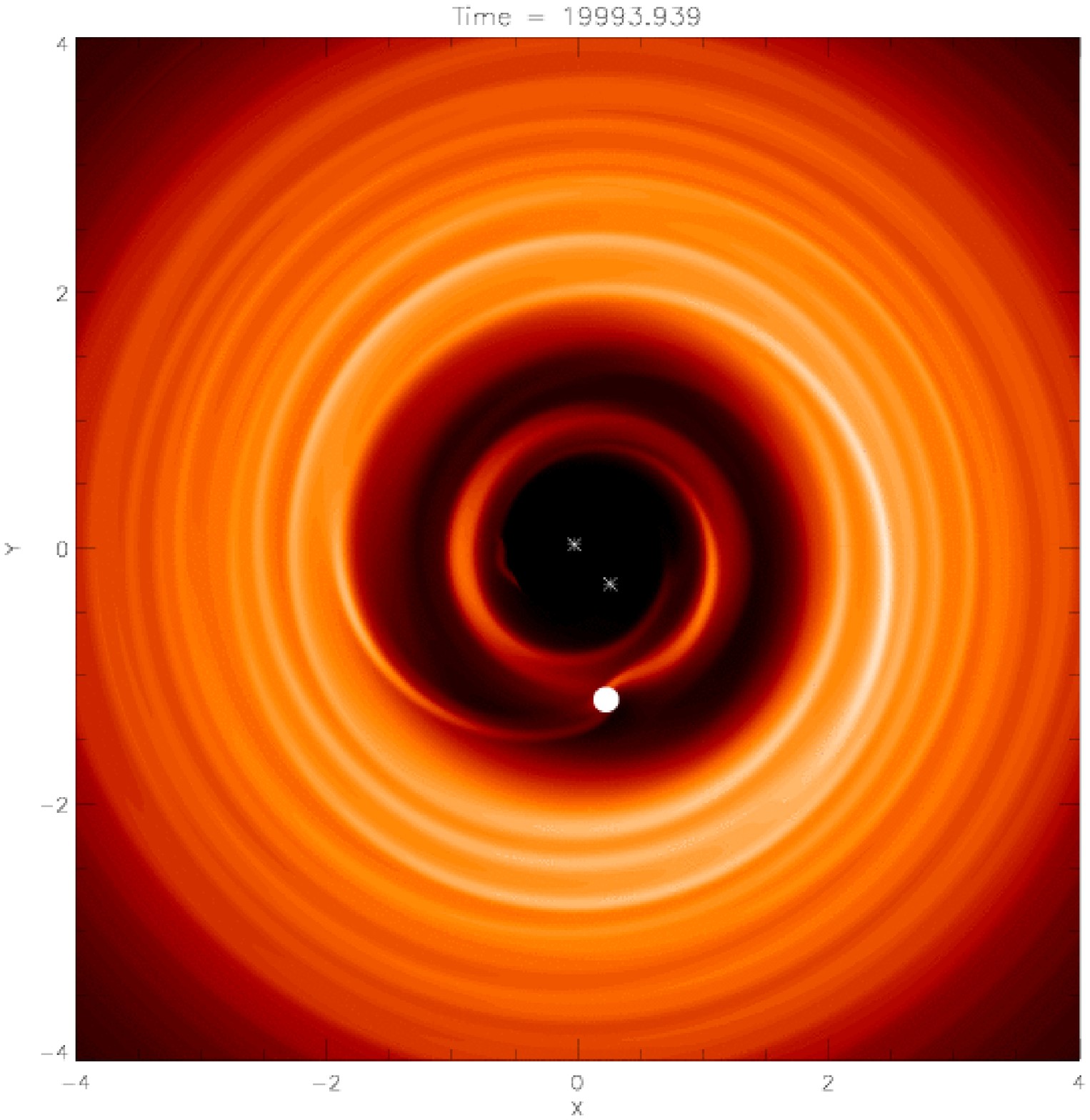}
     \includegraphics[width=6cm]{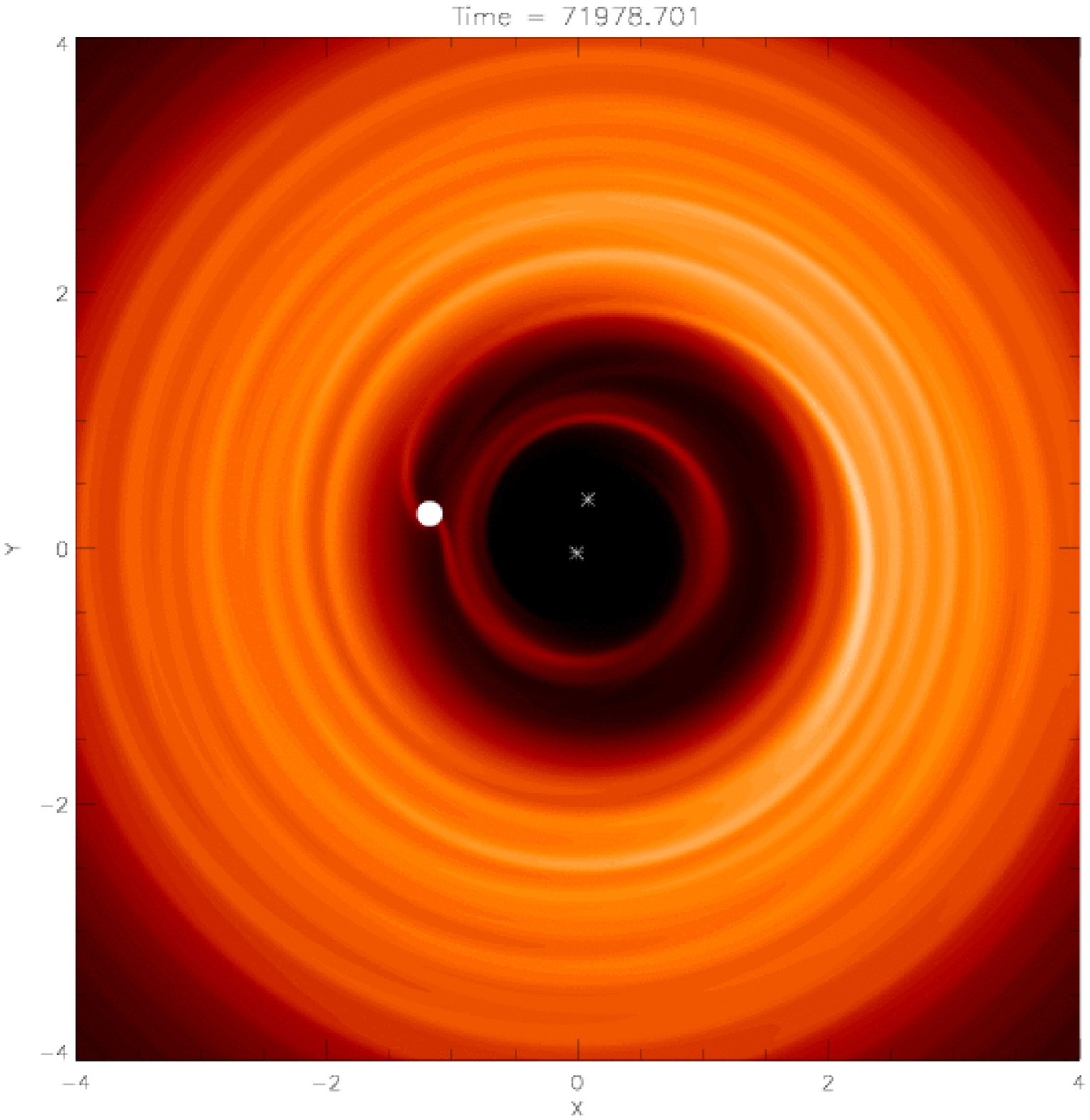}
     \includegraphics[width=6cm]{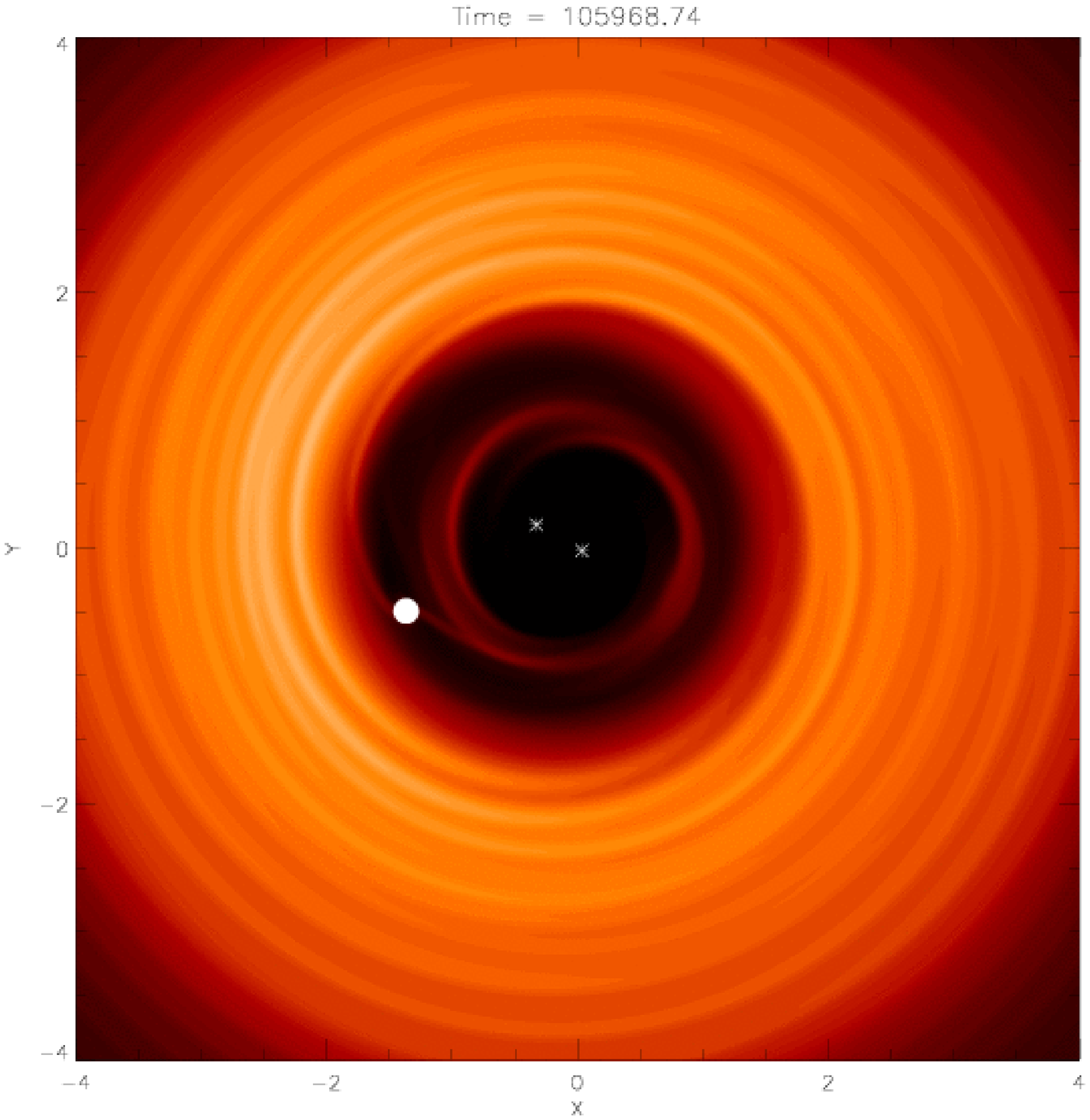}
      \caption{This figure shows, for the model in which $m_p=1\;M_J$
      and $\tacc=\tdyn$,  snapshots of the disc surface density
      at times shown above the plots. Here, the planet is
      represented by a white circle.}
         \label{jupiter2d}
   \end{figure*}

In both simulations, the evolution of the planet subsequent to this 
initial scattering
is quite similar. In the following, we use the
results of the simulation with $\tacc=\tdyn$ to discuss in more details
how the evolution of the system proceeds after this event. 
At $t\sim 10^4 \; P$, 
interaction with the disc and central binary finally results in the 
planet settling into an orbit further out in the disc with eccentricity
$e_p \simeq 0.1$--0.15. The planet now forms a gap in the disc
and begins to migrate inward slowly under type II migration.
The evolution of the disc and planet plus binary system for this 
run is presented in Fig. \ref{jupiter2d}, which shows snapshots of 
the disc surface
density  at different times. The first panel corresponds to a time
shortly before the formation of the 4:1 resonance while the three
other ones display the state of
the system just after the initial scattering. These show clearly that the
inner disc is progressively lost through the inner boundary as a
result of viscous evolution, thereby 
leading to a reduction of the (positive) inner disc torques
and consequently to the inward migration of the planet.\\

\begin{figure*}
   \centering
   \includegraphics[width=\columnwidth]{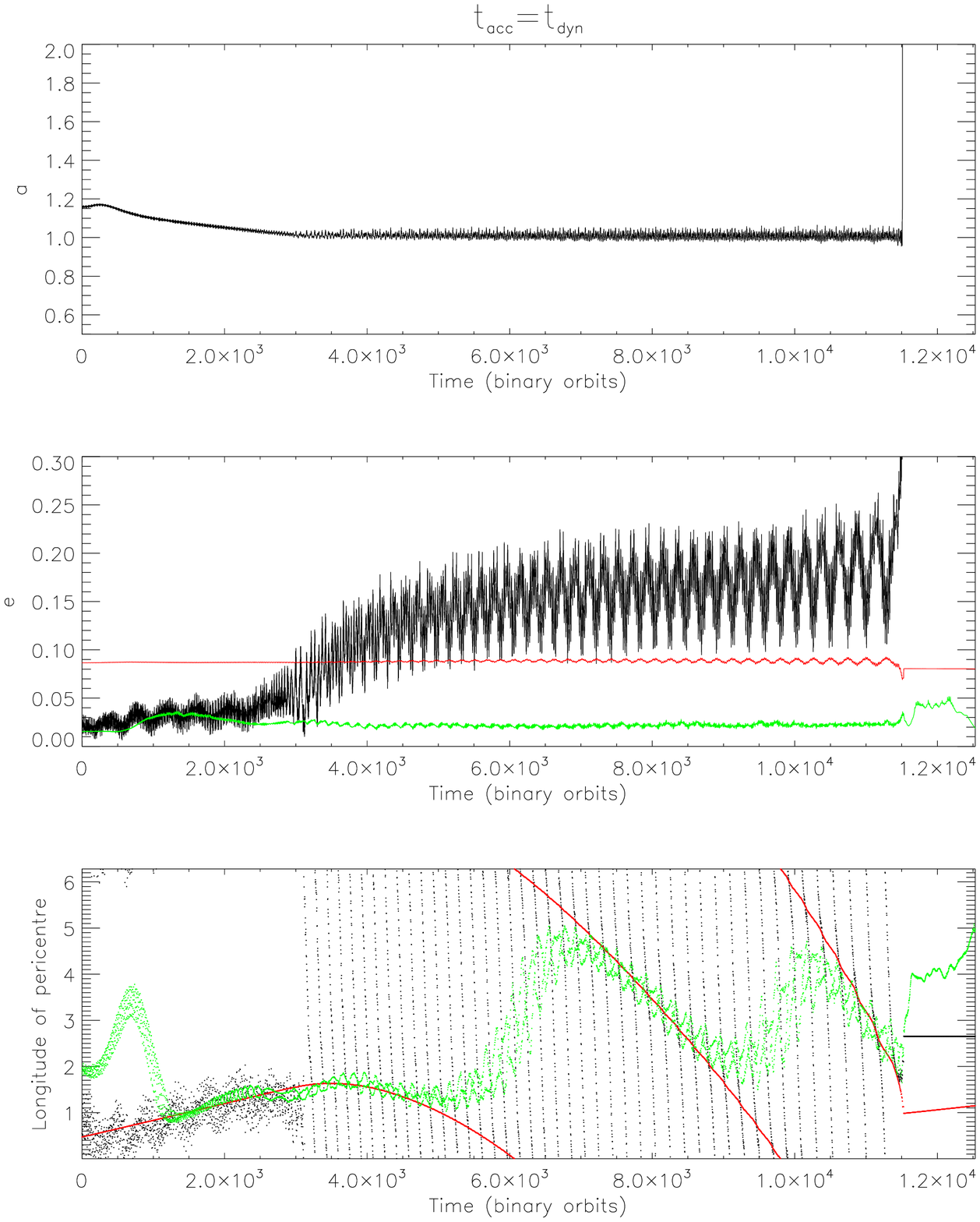}
   \includegraphics[width=\columnwidth]{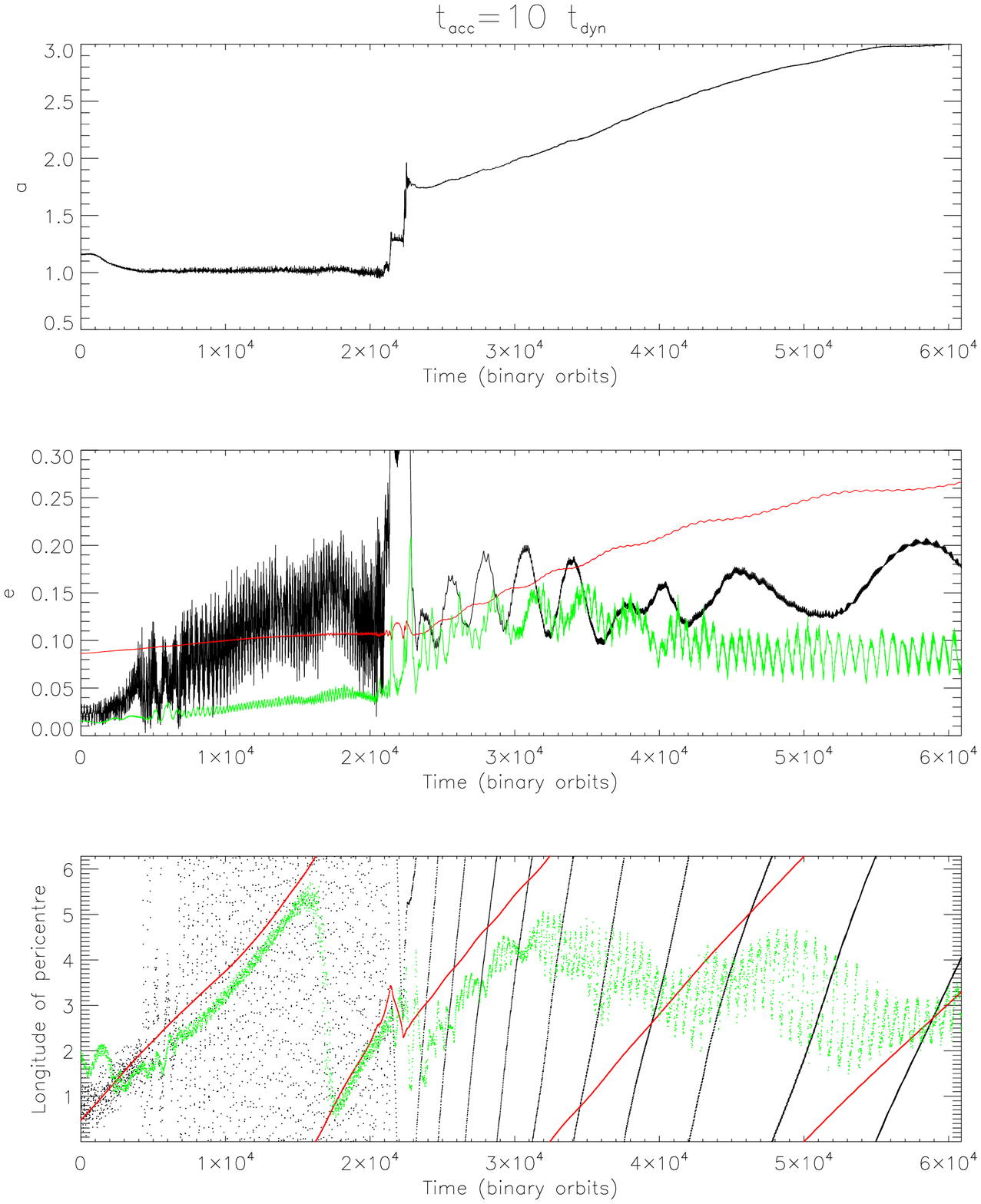}
      \caption{This figure shows the evolution of the system for
      models in which $m_p > 1\;M_J$. Here, the left and right panels correspond
      to $\tacc=\tdyn$ and $\tacc=10\;\tdyn$ respectively. {\it Top:}
      evolution of the semi-major axis of the planet. {\it Middle:}
      evolution of the eccentricities for the planet (black),
      binary (red) and disc (green). {\it Bottom:} evolution of the
      longitudes of pericentre for the planet (black), binary (red)
      and disc (green).}
         \label{giant}
   \end{figure*}
At the end of the run with $\tacc=\tdyn$, the final fate of the planet
is still uncertain despite the very long time scale covered by the
simulation. However, it is likely that the continued inward 
migration of the planet, combined with the growth of the eccentricities 
of both the binary
and planet (see middle panel of Fig. \ref{jupiter}), will result 
in further close encounters and eventually in the planet being 
completely ejected from the
system.
Such an outcome is observed in the calculation with
$\tacc=10\; \tdyn$. In that case, this arises because inward migration 
of the planet causes the temporary formation of the 3:1 resonance 
with the binary at $t\sim 1.35\times 10^5\;P$, which makes the planet 
eccentricity increase up to $e_p\sim 0.15$. This leads to close encounters 
between the planet and the secondary star, resulting in the
scattering and ejection of the planet from the system.\\
The lower panel of Fig. \ref{jupiter} shows the evolution of the disc, binary and planet longitudes of pericentre for both models. Interestingly, we see that here the planet is aligned with neither the disc nor the binary. Therefore, relative to simulations with $m_p=1\; M_S$, the probability of close encounters between the planet and binary is increased, thereby leading to a system which is more likely to become unstable.\\ 
Lastly, we notice that the results of these runs are broadly consistent with previous
hydrodynamical simulations of jupiter-mass planets embedded in
circumbinary discs (Nelson 2003) which indicated  that indeed, trapping into 4:1 resonance
followed by a scattering through a close encounter with
the binary is a possible outcome of such systems.

\subsection{Evolution of giants with $m_p\; > 1\;M_J$}

The evolution of accreting bodies with final masses $m_p > 1\;M_J$ 
is depicted, for both values of the accretion time scale, 
in the upper panel of Fig. \ref{giant}. With respect to the 
simulations presented in Sect. \ref{sec:jup}, here the calculations 
differ in that the planet can continue to accrete gas once its mass 
has attained $m_p=1\; M_J$. Although subsequent evolution may differ, 
this implies that while $m_p < 1 \, M_J$ the evolution of the planet is the same
as the planets considered in Sect. \ref{sec:jup}. 
Therefore, when discussing the results below, we consider only 
the evolution of the system from the time when the planet mass 
has reached  $m_p=1\; M_J$. \\

\begin{figure}
   \centering
   \includegraphics[width=0.95\columnwidth]{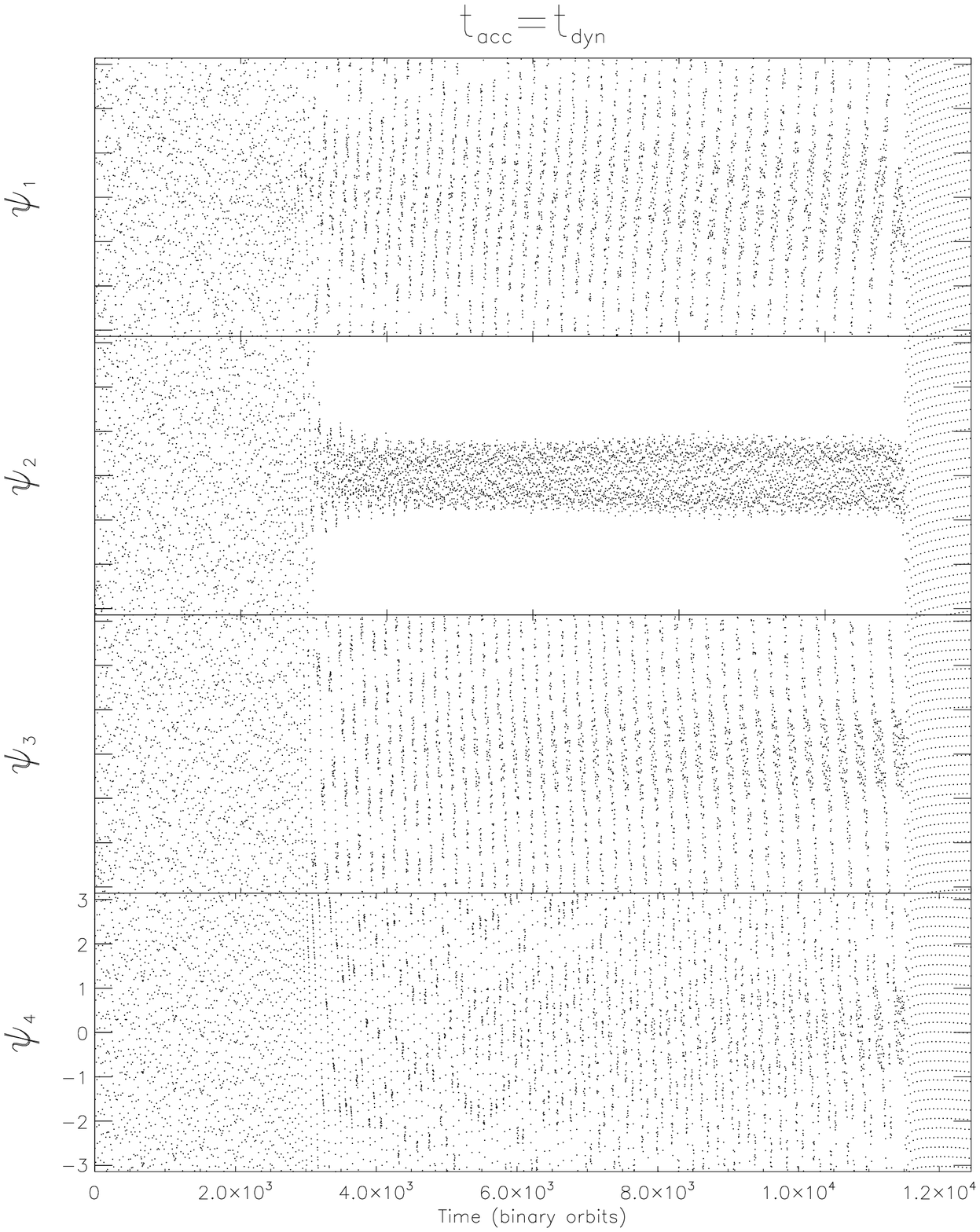}
      \caption{This figure shows, for the simulation with
      $m_p > 1\;M_J$ and $\tacc=\tdyn$, the evolution of the resonant angles  $\psi_1$,  $\psi_2$,  $\psi_3$ and
$\psi_4$ associated  with the 4:1 resonance.}
         \label{giant_angles}
   \end{figure}

For the model with $\tacc=\tdyn$, the planet is in 4:1 resonance 
with the binary when its mass reaches and exceeds that of Jupiter, 
which arises at $t\sim 5.5\times 10^3\;P $ (see Fig.~\ref{mass}). 
Fig.\ref{giant_angles} displays
the evolution of the resonant angles associated with the 4:1 
resonance. Comparing this figure with Fig.\ref{angles}, we see that 
the 4:1 resonance breaks at $t\sim 8.5\times 10^3\; P$ 
with $m_p = 1\; M_J$ whereas it breaks at 
$t\sim 1.15\times 10^4\; P$ in the one with 
$m_p > 1\; M_J$. This supports the idea that the resonant interaction 
is stronger for higher planet masses. Once again, the resonance drives 
the planet eccentricity up to  $e_p\sim 0.2$, until the 
planet undergoes a close encounter with the binary and 
is subsequently ejected from the system at 
$t\sim 1.15\times 10^4\; P$. Here, it is worth noting that 
the planet mass is $m_p\sim 1.7\;M_J$ when this occurs (see Fig.\ref{mass}).\\

\begin{figure*}
   \centering
   \includegraphics[width=6cm]{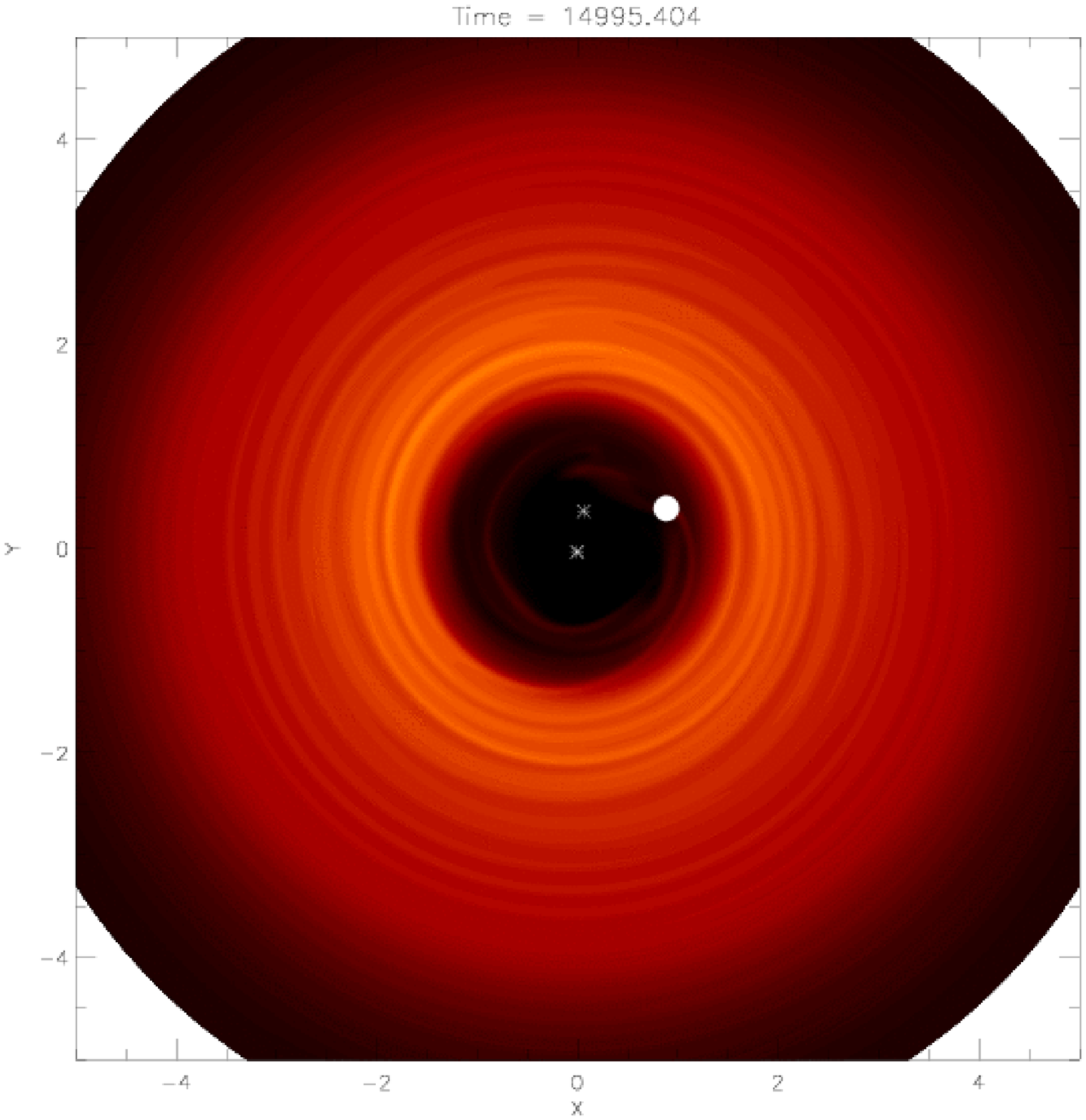}
    \includegraphics[width=6cm]{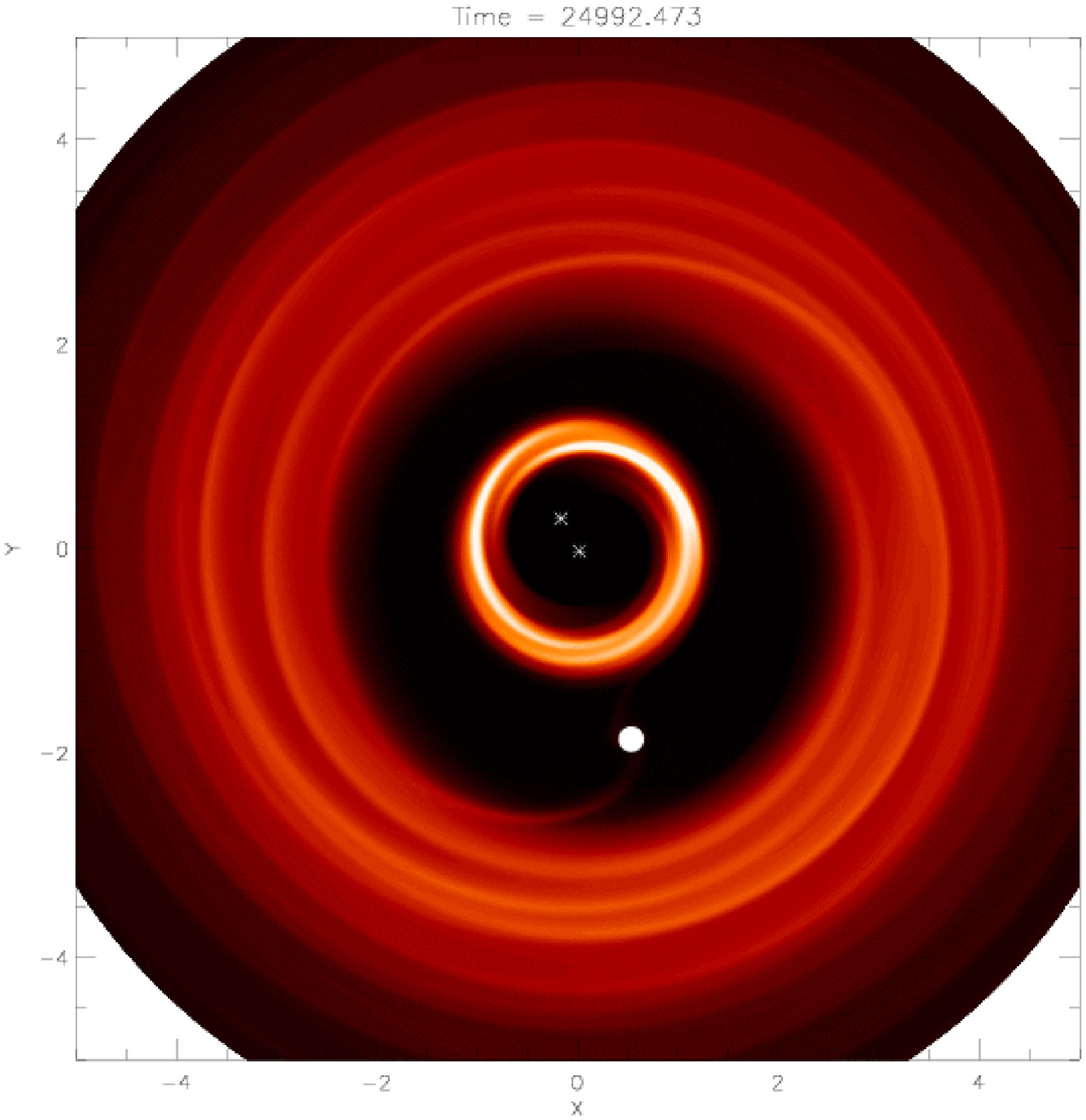}
     \includegraphics[width=6cm]{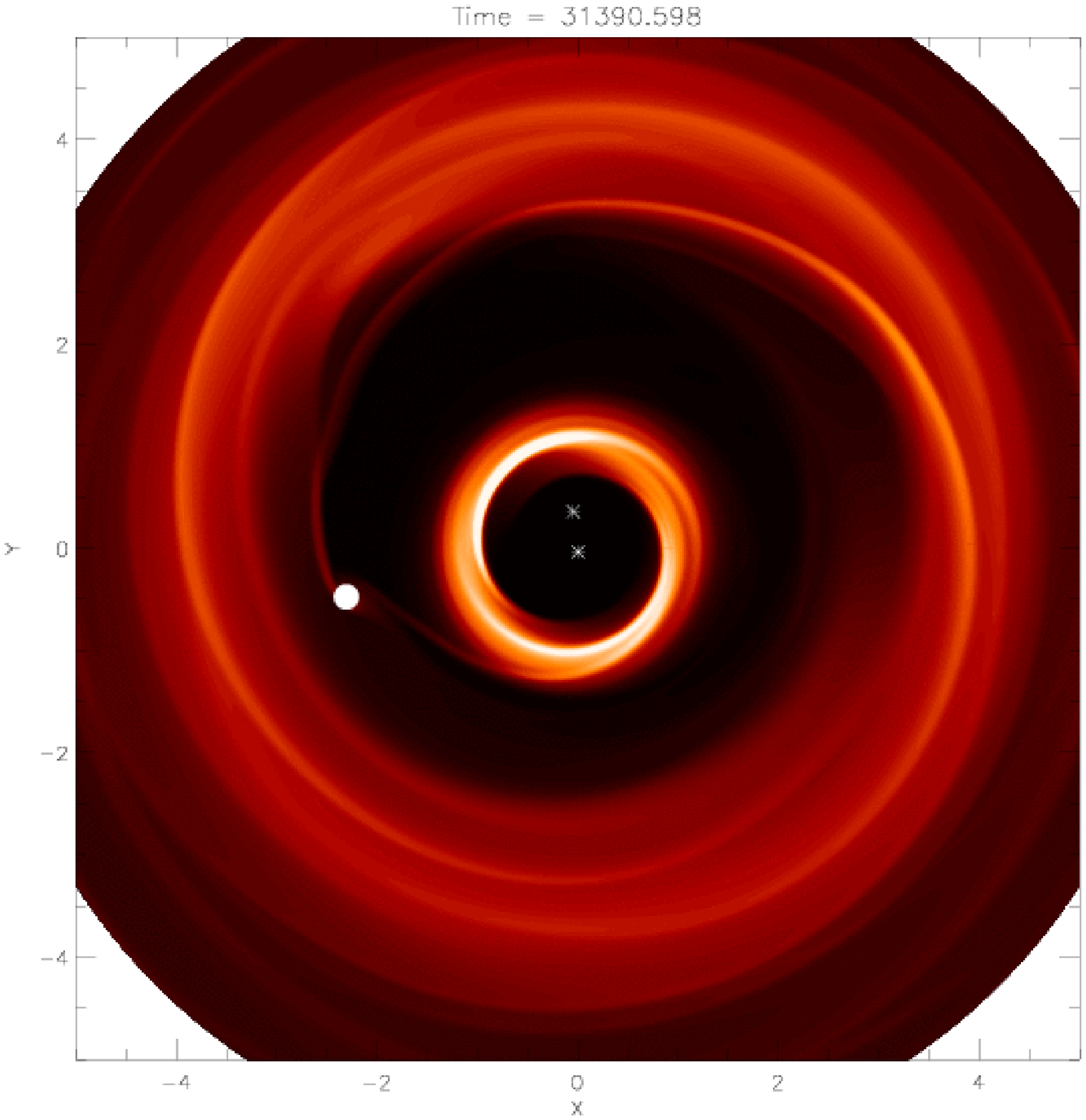}
     \includegraphics[width=6cm]{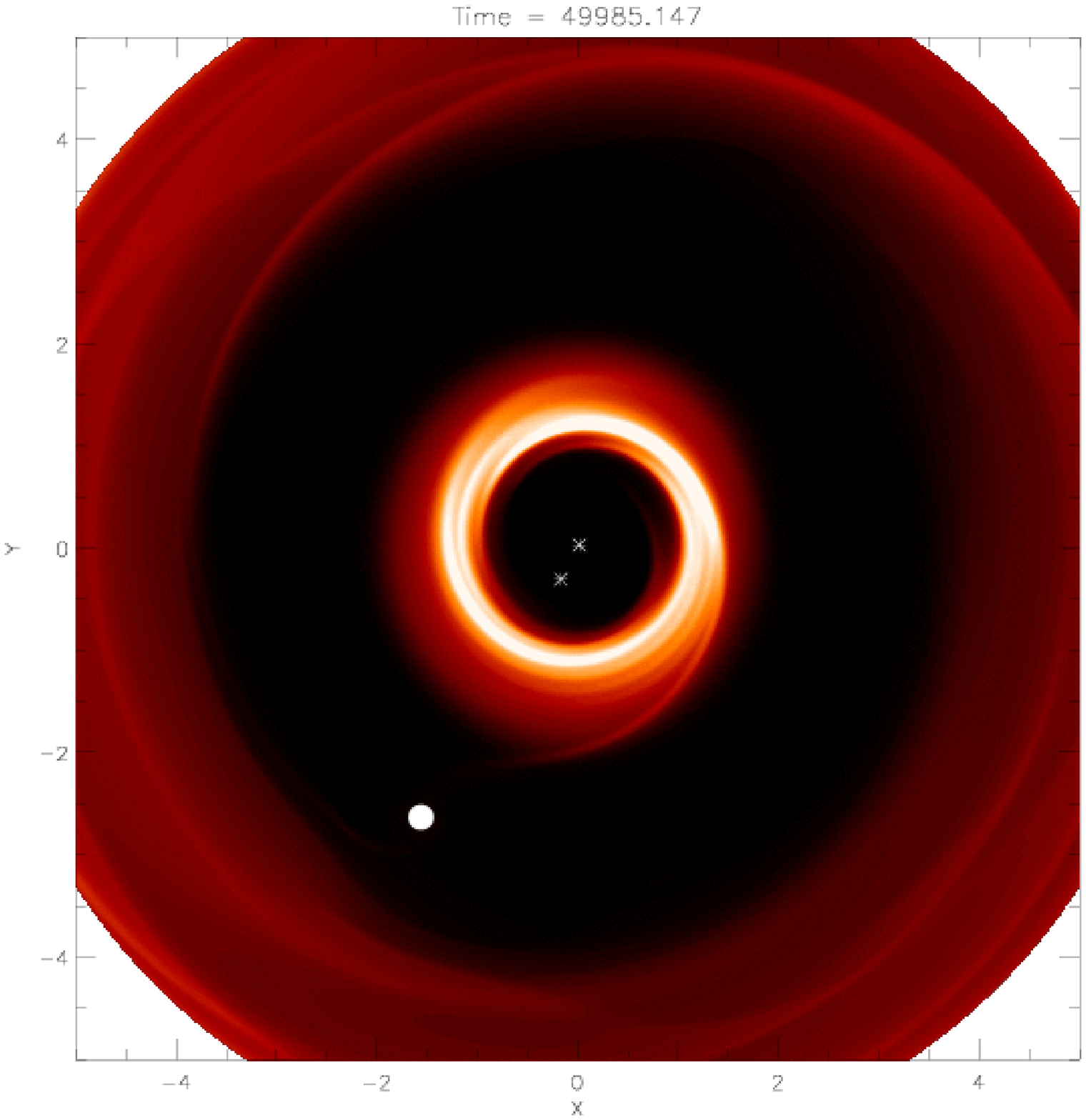}
      \caption{This figure shows, for the model in which $m_p > 1\;M_J$
      and $\tacc=10\; \tdyn$,  snapshots of the disc surface density
      at times shown above the plots. Here, the planet is
      represented by a white circle.}
         \label{giant2d}
   \end{figure*}

\begin{figure}
   \centering
   \includegraphics[width=0.95\columnwidth]{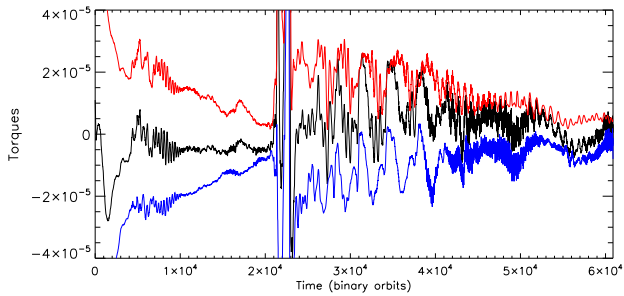}
      \caption{This figure shows, for the model with  $m_p > 1\;M_J$ and $\tacc=10\; \tdyn$, the evolution of the torques exerted by the disc on the planet. The blue (red) line corresponds to the inner (outer) disc torques and the black line corresponds to the net torques exerted on the planet. }
         \label{torques}
   \end{figure}

For the model with $\tacc=10\; \tdyn$, the planet orbits just beyond the location of the 4:1 resonance with the binary when its mass attains $m_p=1\;M_J$, which corresponds to $t\sim 1.4\times 10^4\;P$. From this point in time until
 $t\sim 2.1\times 10^4\;P$, the orbital evolution of the planet is similar to that found in the model with $m_p=1\;M_J$ and $\tacc=10\; \tdyn$ (see Sect. \ref{sec:jup}). Then, the large value of its eccentricity ($e_p\sim 0.13$) causes the planet to undergo a close encounter with the binary and to be scattered out. Interestingly, this scattering leads to the temporary formation of the 6:1 resonance between the planet and the binary at $t\sim 2.15\times 10^4\;P$. As can be seen in the middle panel of Fig.\ref{giant}, which shows the eccentricities of the 
planet, disc and binary, trapping into 6:1 resonance makes 
the planet eccentricity increase to $e_p \ge 0.2$. 
At $t\sim 2.3\times 10^4\;P$, a close encounter between the binary and 
planet occurs, and the planet is consequently scattered out 
on a high eccentricity
orbit with $a_p\sim 1.8$. \\
Just after this scattering, the planet mass has reached $m_p\sim 3\;M_J$ and the eccentricities of the disc and planet have attained $e_d\sim 0.08$ and $e_p\sim 0.1$ respectively. In agreement with Nelson (2003), we find that the interaction
between the eccentric disc and the eccentric planet induces 
outward migration of the latter. The total torque exerted by the disc on the 
planet, as well as the torques due to the disc interior and exterior to $a_p$,
are presented in Fig.\ref {torques}. We see that the time-averaged 
total disc torque is clearly positive from $t\sim  2.3\times 10^4\;P$ onward,
and that the torques oscillate with a period of $\sim 3\times 10^3$ 
binary orbits. Interestingly, these time variations correspond closely
to the precession of the planet relative to that of the disc.
The reason for this is simply that the eccentric disc exerts an
orbit-averaged positive torque
on the eccentric planet when the orbits are misaligned, and exerts a negative
torque when the orbits are aligned. An antialigned configuration causes
the planet to interact with matter whose angular velocity is greater than
that of the planet at apocentre, and this leads to the planet
experiencing a strong positive torque.\\
At later times, the simulation indicates that the outward migration of the 
planet can be sustained over long time scales. There are a number of 
reasons for this.
First, the planet maintains an eccentric orbit and 
experiences a strong positve torque
at apocentre when the disc and planet orbits are antialigned; this positive
torque experienced by the planet implies a negative torque  experienced
by the outer disc material, causing gas to flow through the planet
orbit to form an inner disc. Fig. \ref{giant2d} shows the state of the disc,
planet and binary at different times. The first panel shows the disc
just before the planet enters the 4:1 resonance, and the second panel shows 
the system just after the planet has been scattered outward. The third and 
fourth panels show the increase in inner disc size and the outward migration of
the planet.
During its transition from outer to inner disc, the disc material continues to exert a positive torque on the 
planet, manifested through a corotation torque. Moreover, the existence of
an inner disc provides another source of positive Lindblad torques 
and assists the planet in maintaining outward migration.\\
Examination of the torques exerted by the disc on the planet 
(see Fig. \ref{torques}) shows  
that they weaken from $t\sim 4\times 10^4\; P$. This is due mainly to the
increase in gap size generated by the planet as it grows in mass,
combined with the fact that the disc contains a finite reservoir of
gas in our model. The increase in gap size clearly affects the torques
due to both outer and inner discs. \\

At the end of the run, the planet has migrated to the outer edge of the disc 
and its mass has attained $m_p\sim 5\;M_J$. This suggests that another
avenue for evolution of giant planets whose mass exceeds 1 $M_J$ is long--term
outward migration, in addition to the possibility of being scattered out
of the system as displayed by the runs described previously. Establishing
the ratio of these two outcomes will require a large suite of simulations
with slightly varying initial conditions, task which is beyond the scope
of the paper. It appears from our results, however, that the probability
of scattering and ejection is greater than that of prolonged outward migration.

\section{Summary and conclusion}

In this paper we have presented the results of hydrodynamic simulations aimed 
at studying the formation and evolution of giant planets embedded 
in circumbinary discs.\\
We focused on a model in which a $20\;\mearth$ core initially 
trapped at the edge of the cavity formed by the binary can slowly 
accrete material from the disc. We examined how the final outcome of 
the system depends on the accretion rate onto the 
planet and on the final planet mass attained. For each value of 
the accretion rate considered, we performed three calculations. 
In two of the three simulations, we assumed that accretion stops when 
the mass of the planet has reached either $m_p=1\;M_S$ or 
$m_p=1\;M_J$. In the remaining case, we allowed the planet to accrete gas 
freely from the disc in such a way that its final mass 
was $m_p > 1\;M_J$. The simulations show different outcomes, 
depending on the final mass of the planet:\\
i) In models with $m_p=1\;M_S$, the planet migrates inward until its 
eccentricity becomes large enough for the disc torques exerted 
on the planet to become positive, leading to reversal of migration. 
The planet then enters in a regime of runaway outward migration until 
it passes through the 5:1 resonance with the binary, which halts
the runaway migration. From this time onward the planet first
drifts outward very slowly, and then migrates inward again until
the end of the simulation. Although it is not possible to
run these simulations further, we speculate that this process of
periodic inward and outward migration is repeated, leading to the formation
of a long--term stable circumplanetary system.\\
ii)  In models with final mass $m_p=1\;M_J$, the evolution is 
found to depend 
weakly on the value of the accretion rate. For $\tacc=\tdyn$, 
the planet becomes locked into the 4:1 resonance with the binary until 
it is scattered to a larger radius due to a close encounter 
with the secondary star. For $\tacc=10\;\tdyn$, the planet orbits first in,
and then just outside, 
4:1 resonance. Scattering also arises in this case.
In both cases the planet migrates back in toward the binary system,
and for $\tacc=10\;\tdyn$ the planet undergoes another close encounter
with the central binary which leads to ejection from the system.
The run with $\tacc=\tdyn$ shows slow inward migration of the
planet, such that we are unable to run this model beyond $\sim 1.6 \times 10^5$
binary orbits. We speculate, however, that recurrent episodes of inward
migration followed by close encounters with the central binary
will eventually lead to ejection from the system. An alternative 
outcome, however, is scattering out to a large distance within the disc
where the planet can orbit stably until disc dispersal.
Once again it will require a large suite of simulations
to determine the ratios of these various outcomes, which goes
beyond the scope of this paper. \\
iii) In models with $m_p > 1\;M_J$, the final outcome depends more 
strongly on the value of the accretion rate. In the simulation 
with $\tacc=\tdyn$, trapping into 4:1 resonance with the binary leads 
ultimately to a planet with mass $m_p\sim 1.7\; M_J$ being completely 
ejected from the system. A close encounter between the 
planet and binary is also observed in the calculation with 
$\tacc=10\;\tdyn$. Subsequently however,  the planet is not ejected 
but is rather scattered out to a larger radius on a high eccentric
orbit. After this scattering, the planet migrates outward through its 
interaction with an eccentric disc, until it reaches the outer edge 
of the disc. At the end of the simulation, the planet 
mass has reached $m_p\sim 5\;M_J$. \\
From an observational point of view,  the results of our simulations 
indicate that Saturn-mass planets are probably the best candidates 
to be found in close binary systems. Higher-mass planets are usually 
found to undergo close encounters with the secondary star, 
which raises the possibility that these systems become unstable 
in the course of their evolution.\\
A number of other issues remain to be resolved  when considering the 
early stages of planet formation in circumbinary discs. For example, 
the question of whether or not  planetary cores can grow due 
to planetesimal accretion needs to be examined in more detail. 
An eccentric binary can lead to the formation of an eccentric disc, 
in such a way that planetesimal accretion is prevented except in the 
outer regions of the disc. The ability of planetesimals to accrete,
with care being taken to simulate the structure of the circumbinary 
disc, will be the subject of a future paper. 

Another issue relates to the fact that we have only simulated a
two dimensional system in this paper. The close encounters
experienced by planets with the central binary are likely to
produce significantly inclined orbits, and the resulting disc-planet
interaction is likely to be modified (weakened) by this, leading
to potentially different outcomes to those observed in our 2D runs.
Given the very long evolution times involved, however,
performing full 3D
simulations is beyond current computational capability.
There remains scope, however, for a more approximate 3D treatment 
of this problem.

\begin{acknowledgements}
The simulations performed in this paper were performed on
the QMUL High Performance Computing facility purchased
under the SRIF iniative.
\end{acknowledgements}

\end{document}